\documentclass[preprint2]{proto}
\usepackage{times}
\newcommand{\refs}{\par\noindent\hangindent=1pc\hangafter=1}
\def\phc{\phantom{$\pm$0.00}}

\begin{document}

\title{\textbf{\LARGE Astrophysical Conditions for Planetary Habitability}}

\author {\textbf{\large Manuel G\"udel, Rudolf Dvorak}}
\affil{\small\em University of Vienna}
\author {\textbf{\large Nikolai Erkaev}}
\affil{\small\em Russian Academy of Sciences}
\author {\textbf{\large James Kasting}}
\affil{\small\em Penn State University}
\author {\textbf{\large Maxim Khodachenko, Helmut Lammer}}
\affil{\small\em Austrian Academy of Sciences}
\author {\textbf{\large Elke Pilat-Lohinger}}
\affil{\small\em University of Vienna}
\author {\textbf{\large Heike Rauer}}
\affil{\small\em Deutsches Zentrum f\"ur Luft- und Raumfahrt (DLR) and TU Berlin}
\author {\textbf{\large Ignasi Ribas}}
\affil{\small\em Institut d'Estudis Espacials de Catalunya - CSIC}
\author {\textbf{\large Brian E. Wood}}
\affil{\small\em Naval Research Laboratory}

\begin{abstract}
\baselineskip = 11pt
\leftskip = 0.65in 
\rightskip = 0.65in
\parindent=1pc
{\small With the discovery of hundreds of exoplanets and a potentially huge 
number of Earth-like planets waiting to be discovered, the conditions 
for their habitability have become a focal point in exoplanetary 
research. The classical picture of habitable zones primarily relies 
on the stellar flux allowing liquid water to exist on the surface of 
an Earth-like planet with a suitable atmosphere. However, numerous 
further stellar and planetary properties constrain habitability. 
Apart from "geophysical" processes depending on the internal structure 
and composition of a planet, a complex array of astrophysical factors additionally 
determine habitability. Among these, variable stellar UV, EUV, and 
X-ray radiation, stellar and interplanetary magnetic fields, ionized
winds, and energetic particles control the constitution of upper 
planetary atmospheres and their physical and chemical evolution. 
Short- and long-term stellar variability necessitates full time-dependent 
studies to understand planetary habitability at any point in time. 
Furthermore, dynamical effects in planetary systems and transport 
of water to Earth-like planets set fundamentally important constraints.
We will review these astrophysical conditions for habitability under 
the crucial aspects of the long-term evolution of stellar properties, the
consequent extreme conditions in the early evolutionary phase of planetary 
systems, and the important interplay between properties of the host star 
and its planets.
  \\~\\~\\~}%leave this in to get the correct vertical space after the abstract
 
\end{abstract}

\section{\textbf{INTRODUCTION}}

Discovery of planets around other stars is now well underway. Over 1000 extrasolar planets have been found 
mainly by ground-based radial velocity (RV) measurements and space- and ground-based photometric
transit surveys ({\tt{exoplanets.org}}). In addition,
$\approx 3500$ ``planet candidates'' have been found by Kepler ({\em Batalha et al.}, 2013;  updates
as of November 2013).  Due to instrument sensitivity the first 
planets detected were predominantly gas or ice giants, like Jupiter or Neptune. 
Recent work by {\em Howard} (2013) and {\em Fressin et al.} (2013) show that small transiting 
planets are more numerous than big ones.
Some potentially rocky planets have been identified ({\em L\'eger et al.,} 2009; {\em Batalha et al.,} 
2011, {\em Borucki et al.,} 2013). Several planets with minimum masses consistent with rocky planets 
have been found mostly around cooler stars, including planets in the Habitable Zones 
({\em Pepe et al.} 2011; {\em Anglada-Escud\'e et al.}, 2012; {\em Bonfils 
et al.,} 2013), as these stars are less massive and smaller, and 
hence more easily accelerated by small planets, and show an improved planet/star contrast ratio in the photometric 
transit method. The search is now on for rocky planets around FGKM stars. 
A handful of these have already been found ({\em Buchhave et al.}, 2012; see also website {\tt exoplanets.org}). 
This search is difficult to do by RV because the induced stellar velocities are small. Kepler can 
find close-in rocky planets around FGK stars rather easily compared to planets at larger radii. 
However, small, potentially rocky planets in the habitable zone are difficult to verify, since most 
of the solar-like stars 
are noisier than the Sun, and those identified by Kepler are far away, hence dim 
({\em Gilliland et al.,} 2011).  Nevertheless, the first few transiting planets with radii 
consistent with rocky planets have been recently detected in Habitable Zones 
({\em Borucki et al.}, 2013; {\em Kopparapu}, 2013; {\em Kaltenegger et al.,} 2013).

In this chapter we are concerned with habitable planets. Below, we expand on what we mean by that term. 
A fundamental requirement of habitability, though -- perhaps the only indisputable requirement -- is that 
the planet must have a solid or liquid surface to provide stable pressure-temperature conditions. In 
his book Cosmos, Carl Sagan imagined hypothetical bag-like creatures that could live on Jupiter by 
adjusting their height in the atmosphere, just as some fish can adjust their depth in Earth's oceans using 
air bladders. But one needs to ask how such creatures might possibly evolve. Single-celled organisms could 
not maintain their altitude on a gas or ice giant planet, and hence would eventually be wafted up to the 
very cold (unless it was a hot Jupiter) upper atmosphere or down into the hot interior. Here we assume 
that Sagan's gas giant life forms do not exist, and we focus our attention on rocky planets.
Habitability for extrasolar planets is usually defined by liquid water on the surface of a rocky planet. Other liquids can
be imagined, as discussed below. Furthermore, water may exist also subsurface, as, e.g., expected for Jupiter's 
icy moon Europa, which would make such a  body also potentially habitable. In terms of exoplanets, 
however, such sub-surface life would be very difficult to detect, and we neglect it here. 

% Sect. 2
\section{\textbf{WHAT IS HABITABILITY?}}  

To begin, we must define what constitutes a rocky planet. That definition is easy enough if you can characterize
the planet the way we do for objects in our own Solar System: it must have a solid or liquid surface. But if one knows
only the planet's minimum mass (from RV) or its diameter (from transits), this question becomes more difficult.
Astrophysical theory predicts that planets that grow larger than about 10 Earth masses during the time that the stellar
nebula is still around will capture significant amounts of gas and turn into gas or ice giants ({\em Mizuno et al.}, 1978;
{\em Pollack et al.}, 1996). The theory is not robust, however, because the opacity calculations within the accreting 
atmosphere are complex and did not include such phenomena as collision-induced absorption of IR radiation by molecular hydrogen, 
and because it does not exclude higher-mass planets that complete the process of accretion after the nebular gas has disappeared. 
Similarly, measurement of a planet's diameter via the transit method does not distinguish conclusively between a rocky planet
like Earth or an ice giant like Neptune. Neptune's diameter is about 4 times that of Earth. 
Initial data indicate that planets with radii below $2R_{\rm Earth}$
are rocky. The Kepler team assumes that planets with diameters $> 2R_{\rm Earth}$ are 
ice giants.

Once we are sure that a planet is rocky, the next most fundamental requirement for life as we know it
is that it  should have
access to liquid water. Even though some terrestrial organisms -- those that form spores, for example -- can
persist for long time intervals without water, all terrestrial life forms require liquid water to
metabolize and reproduce. There are good biochemical reasons for this dependence. Most importantly, water
is a highly polar solvent that can dissolve the polar molecules on which carbon-based life depends. That
said, researchers do not completely discount the possibility that alien life forms might have different
biochemistries ({\em Baross et al.}, 2007; {\em National Academy of Sciences}, 2007). Carbon has a more complex chemistry than any
other element, and so most workers agree that alien life would also be carbon-based. Other chain building elements like
silicon and phosphorus exist, but are less energetically favorable than C in this respect. Among alternative solvents, 
even liquid CH$_4$ is considered by some workers to be a possible medium in which life might exist; hence, the interest 
in Saturn's moon  Titan, which has lakes of liquid methane on its surface. Liquid methane is only stable at very low 
temperatures, though, and so any organic chemistry in it may be too slow to create or to power life. We suspect that 
liquid water, or some mixture containing liquid water, is needed to support all forms of life. 

Confining our interest to planets with liquid water is not all that restrictive. Oxygen is the third most
abundant element in the universe, and so many, or most, rocky planets are probably endowed with appreciable 
amounts of water when they form (e.g.,  {\em Raymond et al.}, 2007). In our own Solar System, Mercury and Venus lack liquid water, while Earth and Mars
have it although modern Mars has no liquid water in amounts sufficient to develop life.
The lack of water on the innermost two planets is easily explained, though, as these planets lie
outside the boundaries of the habitable zone -- the region around a star where liquid water can be present on a
planet's surface ({\em Hart}, 1979; {\em Kasting et al.}, 1993). Mars lacks liquid water on its surface today, but
shows evidence that water flowed there earlier in its history. And Mars  may have liquid water
in its subsurface. Mars' internal heat flow is thought to be about 1/3rd that of Earth ({\em Mont\'esi and Zuber},
2003);  hence, depending on the thermal conductivity of the regolith, Mars could have liquid water within a 
few kilometers of the surface.

As astronomers, we are interested in discovering life on extrasolar planets. For the foreseeable
future, at least, such planets can only be studied remotely, by doing spectroscopy on their atmospheres
using big telescopes on the ground or, more likely, in space. For us to detect life on an exoplanet,
that life must be able to modify the planet's atmosphere in such a way that we can detect it, as it has
done here on Earth. Earth's atmosphere contains 21\% O$_2$ and 1.7 ppmv CH$_4$, both of which are produced
almost entirely by organisms. The maintenance of extreme disequilibrium in a planet's atmosphere, and
specifically the coexistence of free O$_2$ with reduced gases such as CH$_4$ or N$_2$O, was suggested many years
ago as the best remote evidence for life ({\em Lederberg}, 1965; {\em Lovelock}, 1965). If we wish to propose a
testable hypothesis about life on other planets, we should therefore restrict our attention to planets
like Earth that have liquid water on their surfaces. This allows for the possibility that
photosynthetic life might flourish there,  greatly increasing productivity ({\em Kharecha et al.},
2005), and possibly producing atmospheric biosignatures that we might one day detect. 

From a practical standpoint, then, this means that we should design our astronomical searches to look
for planets lying within the conventional liquid water habitable zone (HZ) of their parent star. This is 
a necessary, but not sufficient, condition, as factors such as volatile inventories, high-energy radiation, 
magnetospheres, and stellar winds may further constrain habitability, as discussed further in this chapter. The
boundaries of the HZ were estimated by {\em Kasting et al.} (1993) based on 1-D climate modeling
calculations for Earth-like planets. These calculations were done with a cloud-free model (clouds were ``painted'' on the
ground) and they assumed fully saturated tropospheres. 
Conservatively, planets may lose their water
when they lie closer to the star than the ``moist greenhouse'' limit. At this distance, $\approx 0.95$~AU in these
old calculations, a planet's stratosphere becomes wet, and the water is lost by photodissociation
followed by escape of hydrogen to space. A more strict limit of the inner edge of the habitable zone is the so-called 
runaway greenhouse limit. At this limit the entire water ocean of the Earth would reside within the atmosphere due to the
self-enhancing water vapor feedback cycle. The outer edge of the HZ is defined by the ``maximum
greenhouse'' limit, beyond which a CO$_2$-H$_2$O greenhouse is no longer capable of maintaining a warm
surface. This limit is based on the assumption that an Earth-like planet will have volcanoes that emit
CO$_2$ and that CO$_2$ will accumulate in the planet's atmosphere as the surface becomes cold.  But, at some point,
CO$_2$ will begin to condense out of the planet's atmosphere, limiting the range of distances over which this 
feedback process works. In these old
calculations, the maximum greenhouse limit occurs at $\approx 1.67$~AU for our Sun. The ``1st CO$_2$ condensation''
limit of {\em Kasting et al.} (1993) is no longer considered valid,  because CO$_2$ clouds are now thought
to usually warm a planet's surface ({\em Forget and Pierrehumbert}, 1997; {\em Mischna et al.}, 2000). 
{\em Kitzmann et al.} (2013), however, suggest that the warming effect may have been overestimated in some cases.
More optimistic empirical limits on
the HZ are derived from the observation that early Mars appears to lie within it, and Venus could
conceivably have been within it prior to $\approx1$ billion years ago. An even more optimistic outer edge has been
suggested for the HZ by {\em Seager} (2013), based on a calculation by {\em Pierrehumbert and Gaidos} (2011) showing that
super-Earths with dense captured H$_2$ atmospheres could remain habitable out to as far as 10~AU in our Solar System.
Such planets will remain speculative, however, until they are observed, and one would not want to count on their
existence while defining the requirements for a telescope to search for extrasolar life 
({\em Kasting et al.}, 2014).

\begin{figure}[tbp]
  \centerline{
      \includegraphics[scale=0.36]{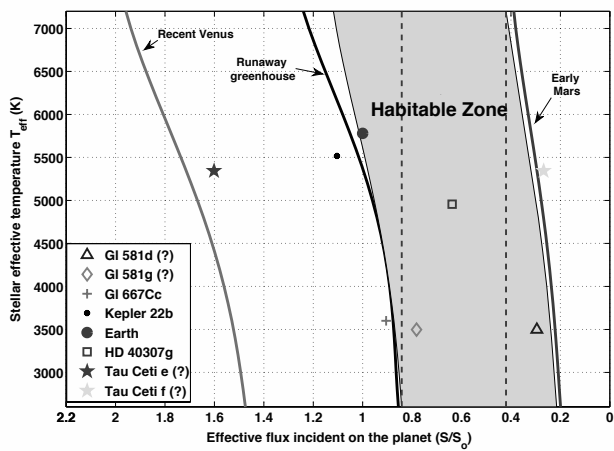}
  }
  \caption{\small Habitable-zone boundaries in incident flux as a function of stellar $T_{\rm eff}$. The
           gray area is delineated by the (conservative) moist greenhouse (inner border) and the maximum 
	   greenhouse (outer border). Somewhat wider HZ are possible, especially related to empirical
	   findings on early Mars and recent Venus. The loci of various exoplanets and the Earth are
	   marked.
           (From {\em Kopparapu et al.}, 2013; \copyright\ AAS. Reproduced with permission.)}
  \label{fig:kopparapu}
\end{figure}

Recently, {\em Kopparapu et al.} (2013) rederived the HZ boundaries using a new 1-D climate model
based on the HITEMP database for H$_2$O. HITEMP differs from the older HITRAN database by including
many more weak absorption lines, including some that extend all the way down to near-UV
wavelengths. They furthermore included a new formulation of the water vapor continuum by {\em Paynter and Ramaswamy} (2011).
This causes the albedo of an H$_2$O-rich atmosphere to be substantially lower than
previously calculated. Consequently, the moist greenhouse limit for the inner edge moves out to
0.99~AU (Fig.~\ref{fig:kopparapu}).  This does not indicate that Earth is actually that close to it, as these calculations
continue to assume a fully saturated troposphere and to omit cloud feedback  ({\em Kitzmann et al.}, 2010; 
{\em Zsom et al.}, 2012). {\em Abe et al.} (2011) and {\em Leconte et al.} (2013)
performed 3D modeling of rocky planets and show that the inner boundary of the HZ strongly depends on the water reservoir of
the planet itself. Water ice may form at the polar caps or on the night side of a tidally locked planet, allowing for liquid
water at some transitional regions. {\em Von Paris et al.} (2010) and {\em Wordsworth et al.} (2010) presented the first 
(1D) studies to simulate the climate of Gl 581d, which lies close to the outer edge of the HZ, finding that the planet 
could be potentially habitable assuming several bars of a CO$_2$-atmosphere. {\em Wordsworth et al.} (2011) studied Gl 
581d with a 3D climate model, also showing that not only the amount atmospheric CO$_2$  determines its habitability but also 
the water reservoir and the resonance between the  orbital and rotational period of the planet. 

The fact that a planet lies within
the HZ of its parent star does not guarantee that it will be habitable. Life depends on other
elements besides C, O, and H and on other compounds besides water. In particular, the elements
N, P, and S all play critical roles in terrestrial life. N is a constituent of both the amino
acids in proteins and the nucleic acids in RNA and DNA. P is needed for the phosphate linkages
between nucleotides in the latter two compounds, and S is needed for some particular amino
acids. The compounds formed by these elements are termed volatiles because they are typically
gases or liquids at room temperature. Habitable planets therefore require sources of volatiles.
Earth's volatiles are mostly thought to have originated from the asteroid belt or beyond
({\em Morbidelli et al.}, 2000; {\em Raymond et al.}, 2004); hence, modes of planetary accretion must be considered. To
hold onto its water, a planet needs an atmosphere in addition to H$_2$O vapor. Planets that are too
small, like Mars, are thought to lose their atmospheres by various thermal and non-thermal
processes, as discussed later in this chapter. Planetary magnetic fields may also be important.
Mars, which lacks an intrinsic magnetic field, has had much of its atmosphere stripped by direct
interaction with the solar wind. Venus, on the other hand, also lacks an intrinsic magnetic
field; yet it has successfully held onto its dense CO$_2$ atmosphere, suggesting that planetary
size is important. Atmospheric composition may also be important, as a less CO$_2$-rich atmosphere
on Venus would have been hotter and more extended in its upper parts and may not have been as
well retained. In the following sections of this chapter, we discuss in more detail various
factors that contribute to planetary habitability. We focus on {\it astrophysical} conditions, 
while geophysical factors (such as internal heat sources, plate tectonics, land mass fraction, 
outgassing, volcanism or magnetic dynamos), lower-atmosphere issues (climate, winds, clouds), 
or biological conditions (damaging radiation doses, extreme biological environments) are
not considered here.

% Sect. 3
\section{\textbf{DYNAMICS AND WATER TRANSPORT}}  

\bigskip
\noindent
\textbf{3.1 Dynamics and Stability}
\bigskip

The evolution of a biosphere is a very gradual process occurring over long time intervals.
The long-term stability of a terrestrial planet moving in the HZ is therefore 
certainly one of the basic requirements for the formation of life on such a planet. 
The eccentricity  of a planetary orbit is a crucial factor in orbital dynamics 
because only a sufficiently small value guarantees that the planet remains
within the HZ. However, to assure appropriate conditions 
for habitability, the dynamics of the whole planetary system
matter. For extensive reviews, we refer the reader to the chapters by
{\em Raymond et al.}, {\em Baruteau et al.}, {\em Benz et al.}, and {\em Davies et al.}
in this volume. We focus on selected issues directly relevant for habitable-zone
planets here. 

Observations of planetary systems and numerical
studies indicate that planets are not born where we find them
today. Simulations of the late stage of formation of the outer solar
system show that the giant planets arrived at their current configuration
through a phase of orbital instability. The so-called {\em Nice Models}
(e.g., {\em Tsiganis et al.}, 2005, {\em Morbidelli et al.}, 2007) provide the 
best-matching results for the final configuration of the outer solar system. 
These simulations start with a more compact system of the four giant 
planets (in a multi-resonant configuration in the latest version); it
then experiences slight changes in the 
orbits due to planetesimal-driven migration leading to a resonance crossing of
Saturn. This crossing causes an unstable phase for the giant planets
with faster migration toward the final orbits. While during
planetesimal-driven migration dynamical friction dampens orbit eccentricities, 
the resonance crossing will increase eccentricities and the orbits 
will diverge from each other ({\em Chiang}, 2003; {\em Tsiganis et al.}, 2005). 

Since Jupiter and Saturn exert a strong influence on the inner
solar system through secular perturbations, a change of the orbital separation
of the two giant planets will modify the secular frequencies of their orbits.
For example, {\em Pilat-Lohinger et al.}\ (2008) studied the secular influence of
Jupiter and Saturn on the planetary motion in the HZ of a G2V star for various
separations of the two giants. While  Jupiter was fixed at its actual orbit 
5.2~AU, Saturn's initial semi-major axis was varied between 
8.0 and 11.0~AU. 
\begin{figure}[tbp]
  \centerline{
      \includegraphics[scale=0.75]{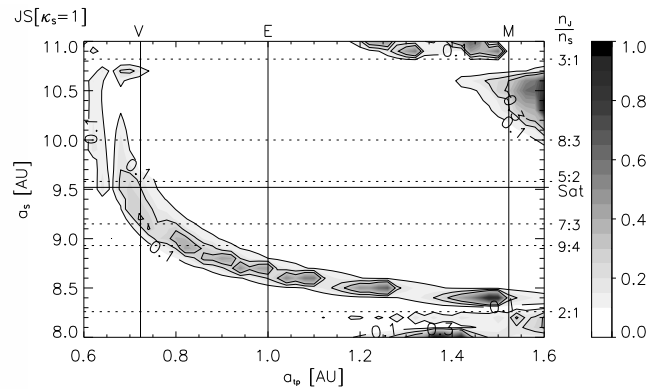}
  }
  \caption{\small Maximum-eccentricity map for test planets under the
    influence of Jupiter and Saturn. The x axis shows the initial semi-major
    axis of the test planet and the y axis indicates the variation of the
    semi-major axis of Saturn. Vertical solid lines indicate the positions
    of Venus (V), Earth (E) and Mars (M) and the horizontal solid line shows the
    actual position of Saturn (Sat). The dashed horizontal lines represent the
    mean motion resonances (MMRs) of Jupiter and Saturn. The grey scale 
    indicates  maximum eccentricities of the test planets.
  (From {\em Pilat-Lohinger et al.}, 2008; \copyright\ AAS. Reproduced with permission.)}
  \label{fig:JupSat}
\end{figure}

Fig.~\ref{fig:JupSat} gives a global view of the perturbations of various
Jupiter-Saturn configurations on the HZ of a Sun-type star. We can clearly see
that most of the area in the $a_{Tp}$-$a_{s}$ plot (semi-major axes for the 
terrestrial planet and Saturn, respectively) is not perturbed by the two
giant planets, but note the arched band showing higher
maximum eccentricities for test planets. This is due to the
secular frequency associated with the precession of the perihelion of Jupiter.
Special attention should be paid to the location of
Earth's orbit, where strong variations of the eccentricity (up to $\approx$0.6)
can be observed for semi-major axes of Saturn between 8.5 and 8.8~AU. We thus see
that the dynamical evolution of giant planets is of utmost relevance
for habitability conditions on Earth-like planets.

The huge diversity of planetary systems differing significantly from the
architecture of our solar system suggests very different evolutionary tracks where
processes like planet migration, planet-planet scattering, resonant
trapping, mutual collisions or hyperbolic ejections influence the final constitution
of a planetary system. Observations reveal large eccentricities
for many exoplanets, probably indicating phases of strong interactions.
  
Since all planets in our solar system move in nearly circular orbits, it appears
that its instability phase was moderate, probably enabling the successful
formation and evolution of life on Earth. It may then be questionable 
whether planetary systems harboring a giant planet in an eccentric orbit can 
provide suitable conditions for habitability at all.

On a further note, recalling that most of the stars in our galaxy are members of binary 
or multiple star systems, additional gravitational perturbations from the binary
interactions should also be taken into account for the assessment of habitability,
as shown in detail by {\em Eggl et al.} (2012), {\em Kane and Hinkel} (2013), 
{\em Kaltenegger and Haghighipour} (2013), and  {\em Haghighipour and Kaltenegger} (2013).

\bigskip
\noindent
\textbf{3.2 Water Transport}
\bigskip

Where did the essential water on Earth in the amount of 0.02 percent then come from? 
Simulations show (e.g., {\em O'Brien et al.}, 2006)
that our terrestrial planets likely formed out of material located in a ring around 
their present locations, i.e., they formed in the region of the habitable zone.
However, both  meteoritic evidence as well as evolutionary disk models show that
the disk in this region was too warm to allow for the presence of sufficient amounts
of water in solids.  Therefore,  water (ice) delivery should have happened later; the 
sources were unlikely to be comets, because statistically their collision probability 
with the Earth is too low ({\em Morbidelli}, 2013). Current thinking is 
that the chondritic material from the outer asteroid belt is primarily responsible 
for the Earth's water content. Water from chondritic asteroid material also 
- in contrast to many comets - on average matches the Earth ocean water 
D/H ratio. For a summary on these aspects, we also refer to the chapter 
by {\em van Dishoeck et al.} in this volume.

The growth of the planetesimals to protoplanets and finally planets is closely
connected with the collisions of the bodies in the early evolution of a
planetary system. Such collisions may lead to
the enrichment of water, depending on the origin of the colliding bodies.
Various simulations (e.g., {\em Raymond et al.}, 2004, 2009; {\em O'Brien et al.}, 2006;
see chapter by {\em Raymond et al.} in this volume) 
produce solar systems akin to ours, but aiming at reproducing properties of our own
solar system in detail has repeatedly led to results  contradicting one or several 
properties (masses of inner planets,
eccentricities, water content of inner planets, asteroid belt architecture; 
{\em Raymond et al.}, 2009). The simulations, however, very sensitively depend on 
the initial conditions and the distribution of the planetesimals. 
Growth of terrestrial planets via collisions (with bodies 
from the dry inner regions, and also the water-rich outer regions of the asteroid 
belt) is often thought to take place when the disk gas has disappeared 
and only massive solid bodies remain. The mutual perturbations 
between these planetary embryos lead to orbital crossings and 
collisions; additionally, giant planets such as Jupiter trigger collisions.

The general assumption now is that about $10\%$ of the mass of the Earth
accreted from beyond 2.5~AU, suggesting that the composition of the relevant
planetary  embryos is close to the composition of carbonaceous chondrites which
contain of order 5--10\%  of water (e.g., {\em Lunine et al.}, 2011). 

\begin{figure}[h!]
\begin{center}     
 \includegraphics[width=3.3in,angle=0]{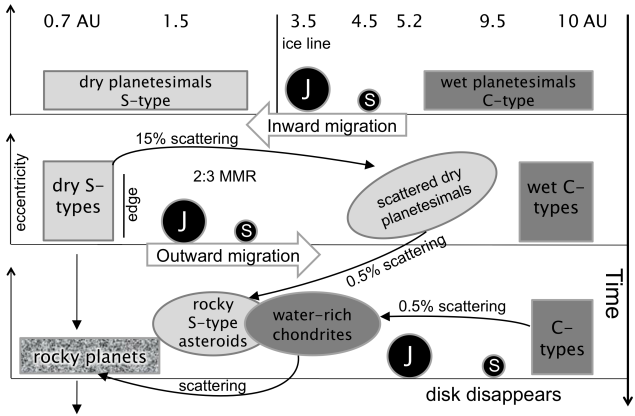} 
% \vspace*{-1.0 cm}                      
 \caption{\small Schematic view of the formation of planets in the habitable zone of
   the solar system according to the ``Grand Tack'' model (after {\em Walsh et al.}, 2011; {\em Morbidelli}, 2013).}
   \label{fig:tack}
\end{center}
   \vskip -0.3truecm
\end{figure}

{\em Walsh et al.} (2011) developed a new model that seems to explain several
solar-system features:
A planetesimal disk still embedded in gas is present inside 3 AU (i.e., volatile-poor
S-types) as well as between and outside the gas giants (water-rich C-types).  
Jupiter migrated inward when Saturn was still accumulating mass; Saturn also started 
to migrate inward because of scattering of rocky planetesimals 
out from the inner part of the belt. The two planets eventually fell into 
the 2:3 mean motion resonance, with Jupiter located at 1.5~AU; meanwhile, the giants had scattered 
S-type planetesimals outward and confined the inner planetesimal disk to about 1-1.5~AU.
Now, at resonance, the inward migration stopped and both giants
started to migrate outwards again, up to the moment when the disk gas had dissipated. 
During the outward motion, Jupiter and Saturn scattered predominantly dry respectively 
wet planetesimals back into the inner solar system, forming the asteroid belt with its
characteristic gradient of volatiles. Further scattering and collision of 
the forming terrestrial planets with the water-rich planetesimals eventually
accumulated water on the inner planets.
This ``Grand Tack'' scenario is depicted in Fig.~\ref{fig:tack} (sketch after the 
original articles by {\em Walsh et al.}, 2011 and {\em Morbidelli,} 2013).
Its end point could be the starting point for 
the ``Nice Model'' mentioned in Sect. 3.1, organizing the architecture of the 
present-day outer solar system and finally inducing the Late Heavy Bombardment 
after some 800~Myrs. The ``Grand Tack'' model successfully explains several features relevant
for habitability of the inner solar-system planets (eccentricities, planetary masses, 
water content).

It is important that the collisions building up the water content on Earth
occurred with velocities smaller than the escape velocity from the planetesimals or 
protoplanets, in order to ensure that  complete merging of the bodies 
occurs ({\em Stewart and Leinhardt}, 2012; {\em Dvorak et al.}, 2012, {\em Maindl \& Dvorak}, 
2014; {\em Maindl et al.}, 2013). 

Obviously, the scenario for the water delivery to Earth is still
incomplete and depends highly on the initial conditions one is using.

% Sect. 4
\section{\textbf{THE HOST STAR IN TIME}}  

The conditions on the surface and in the atmospheres of terrestrial planets are
 controlled by the radiative and particle output of the host star; as this
output varies on time scales of minutes to billions of years, conditions in planetary 
atmospheres are temporarily or permanently modified. 

The photospheric radiative output is a consequence of the nuclear energy production in
the star's interior; it dominates the optical and partly the near-ultraviolet (NUV) spectrum.
The star's effective temperature and its luminosity follow
from the laws of stellar structure and consequent nuclear reactions in the core, which 
gradually vary in time. 

In contrast, radiation in the ultraviolet (UV, 2100-3500~\AA), the far-ultraviolet (FUV, 912--2100~\AA),
the extreme ultraviolet (EUV, 100--912~\AA), and the X-ray ($<$~100~\AA) range is a consequence
of (often violent) energy release in the stellar atmosphere due to annihilation of 
unstable magnetic fields. Magnetic activity is expressed in cool photospheric spots, 
chromospheric plage emitting in the ultraviolet, and hot (million-degree) coronal plasma 
in ``magnetic loops'' with strong X-ray radiation; also, highly-energetic, accelerated 
particles, shock fronts from mass ejections, short-term energetic flares and perhaps
also the solar wind indicate the operation of magnetic energy release. Magnetic fields
are ultimately generated in a magnetic dynamo in the stellar interior, driven by the
interaction between convection and differential rotation. The short-wavelength
stellar output is therefore a function of stellar rotation (depending on stellar age)
and of the stellar spectral type (determining the depth of the convection zone, itself 
also somewhat dependent on age).

A late-type star's rotation period, $P$, steadily declines during its main-sequence life (see chapter 
by {\em Bouvier et al.} in this volume).
This is  consequence of the stellar magnetic field permeating the ionized stellar wind
which transports away angular momentum. The declining rotation period itself
feeds back to the dynamo, weakening the production of magnetic activity. As a consequence,
$P$ not only declines in time but it converges, after a few
hundred million years for a solar analog, to a value  dependent on age but independent
on the initial Zero-Age Main Sequence (ZAMS) rotation rate. Magnetic activity and the related XUV 
(FUV, EUV, and X-ray) output
therefore also become a function of stellar age. A generalized formula has been derived by 
{\em Mamajek and Hillenbrand} (2008) namely $P=0.407([B-V]_0 - 0.495)^{0.325}~t_6^{0.566}~{\rm [d]}$ 
where $[B-V]_0$ is the intrinsic stellar $B-V$ color and $t_6$ is the stellar age in Myr.
For younger stars, the rotation period also depends on the presence of a circumstellar 
disk (around pre-main sequence stars) leading to ``disk-locked'', relatively slow
rotation, and the history of disk dispersal before the arrival on the main sequence (MS),
leading to a wide range of ZAMS rotation periods ({\em Soderblom et al.}, 1993).

\bigskip
\noindent
\textbf{4.1 The Evolution of Stellar Optical Radiation}
\bigskip

Before arriving on the MS, a star's luminosity first decreases when
it moves down the Hayashi track as a T Tauri star; in that phase, the Sun was
several times brighter than today, although its effective temperature was lower ($T \approx 4500$~K)
and its radius consequently larger ($\approx 2-3 R_{\odot}$). Before reaching the
ZAMS, solar-like stars reach a minimum bolometric luminosity, $L_{\rm bol}$, which for a solar-mass 
star is as low as $\approx$0.45$L_{\rm bol, \odot}$ (at an age of $\approx 13$~Myr; {\em Baraffe et al.}, 1998).

The Sun's bolometric luminosity has steadily increased since its arrival on the
ZAMS, starting from $\approx 70$\% of today's luminosity. At 
the same time, the effective temperature also increased somewhat ({\em Sackmann \& Boothroyd},
2003). This increase is a consequence of the evolution of nuclear reactions in the
core of the star. Similar trends hold for other cool main-sequence stars.

During the later MS evolution, the Sun's brightness will increase up to $\approx 3L_{\odot}$ at an
age of 10~Gyr. The significant change in $L_{\rm bol}$ implies that the classical habitable zone around the star
moves outward during this evolution. {\em Kasting et al.} (1993) derived the HZ-width for
the entire main-sequence evolution of the Sun (and cooler stars), defining the ``continuously habitable
zone''. The latter becomes narrower as the inner border moves outward while - under the
assumption that planets moving into the HZ at a later time do not evolve into habitable planets -
the outer border remains the same. Therefore, the total width of the HZ narrows as the star 
evolves. The Earth will in fact be inside the inner HZ radius at an age of $\approx 7$~Gyr, i.e., only
2.4~Gyr in the future (for the most pessimistic models) and so it will lose its water. This effect is less 
serious for later-type stars for the same time interval because of their slower evolution.

\bigskip
\noindent
\textbf{4.2 Evolution of Stellar Ultraviolet Radiation}
\bigskip

The stellar FUV-NUV flux ($91.2 \le \lambda \le 350$~nm) covers the transition
from the emission spectrum of the hot chromospheric/transition region gas, at typical
temperatures of $10^4-10^5$~K, to the rising photospheric emission, dominating
above $\sim$200~nm. This wavelength range contains the strongest emission line,
H Ly$\alpha$, which is responsible for over 50\% of the stellar high-energy (XUV)
flux. The UV domain is especially relevant in the context of planetary
atmospheres since UV photons trigger a variety of photochemical reactions
(e.g., {\em Hu et al.}, 2012; {\em Yung and DeMore}, 1999; Sect. 5.1 below). 
Important molecules such as H$_2$O, CO$_2$, CH$_4$,
CO, NH$_3$, have photodissociation cross sections that become significant below
200~nm. Also, the abundance of the biosignature molecule O$_3$ is mainly controlled by 
photochemical reactions, hence understanding the interaction of the stellar radiation 
with the atmospheric chemistry is of great importance for the determination of possible 
atmospheric biosignatures. A reliable calculation of the 
photodissociation rates and their time evolution is very complex due to the interplay between the
strongly decreasing cross sections and the rapidly rising stellar flux toward longer wavelengths. 
For example, the flux of solar-like stars increases by more than 2 orders of
magnitude from 150 to 200~nm and the photodissociation cross section of, e.g.,
H$_2$O, decreases by over 3 orders of magnitude in the same interval.

The FUV-NUV flux from low-mass stars is variable over many time scales,
including hours (flares), days (rotation period), years (activity cycles) and
Gyr (rotational spin-down). We focus here on the long-term evolution, which is 
comparable to the nuclear time scale. The UV photospheric flux (i.e., $\lambda
\ga 200$~nm) rises with time as the star evolves off the ZAMS and increases its
radius and effective temperature. For the case of the Sun, the overall Main
Sequence UV flux increase ranges within 30\%--200\% depending on the
wavelength, as illustrated by {\em Claire et al.} (2012) on the basis of Kurucz
atmosphere models. 

\begin{figure}[tbp]
  \centerline{
      \includegraphics[scale=0.72]{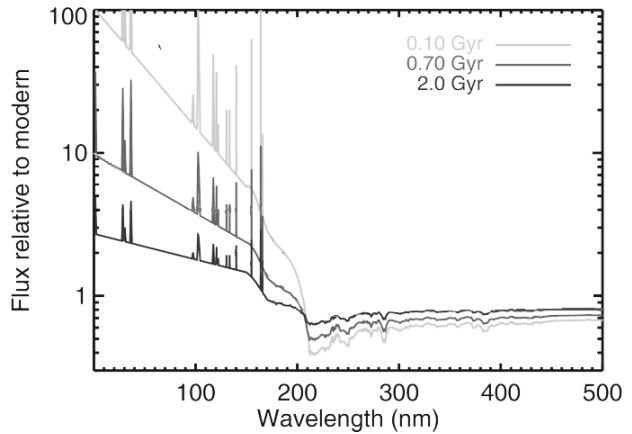}
  }
  \caption{\small Flux ratios relative to the modern (present-day) Sun at 1 AU 
  covering the UV range of solar-like stars of different ages. Both continuum 
  and line fluxes are included using the procedure in {\em Claire et al.} (2012). 
  (Adapted from {\em Claire et al.} 2012; \copyright\ AAS. Reproduced with permission.)}\label{UVflux}
\end{figure}

The chromospheric and transition region emission is much more difficult to
describe, as it depends on the evolution of the magnetic properties of the
star. A successful technique employed to investigate the long-term evolution of
stellar activity is to identify stars that can act as proxies for different
ages (e.g., {\em Ribas et al.}, 2005). In the UV this is especially complicated by the
difficulty in finding stars that are close enough to an evolutionary track and
by the need to use space facilities (FUSE, GALEX, IUE, HST) that have only been
able to provide high-quality data for a limited number of nearby objects. The
best example of the application of the proxy technique is the star
$\kappa^1$~Cet, which is a close match to the Sun at an age of $\sim$0.6~Gyr, as
presented by {\em Ribas et al.} (2010). {\em Claire et al.} (2012) generalized the study to
the entire main sequence evolution of solar-like stars, resulting in 
approximate flux multipliers as a function of wavelength that can be applied to
the current solar flux to calculate the spectral irradiance at different ages.
Figure~\ref{UVflux}  plots the flux multipliers.
Better constraints to the UV flux evolution of solar-like stars will
follow from observations with new instrumentation ({\em Linsky et al.}, 2012).

The flux multipliers of {\em Claire et al.} (2012) are appropriate for the Sun and
solar-like stars but may not be adequate for stars of lower mass. Measurements
of UV fluxes of a handful of M dwarfs are available ({\em France et al.},
2013, and references therein). {\em France et al.} (2013) show that the flux in the FUV band 
relative to the NUV band for most M dwarfs is 3 orders of magnitude larger
than the solar value, and this has strong photochemical implications. But, in
spite of these measurements, we still lack information on the time evolution of
such fluxes, the main reason being the difficulty in estimating stellar ages.
The development of new alternative age determination methods for MS
K and M dwarfs ({\em Garc\'es et al.}, 2011) should help to define the necessary 
UV flux-time relationships.

The determination of the H Ly$\alpha$ line flux deserves special attention.
Precise measurement of the H Ly$\alpha$ line is extremely challenging due to
contamination from Earth geocoronal emission and the difficulty of
correcting interstellar and astrospheric absorption ({\em Wood et al.}, 2005). 
{\em Linsky et al.} (2013) present a compilation of HST/STIS observations of a relatively
large sample of stars from F to M spectral types, with Ly$\alpha$ fluxes and
correlations with other emission line fluxes. When normalizing fluxes to the
stellar habitable zone, {\em Linsky et al.} show a clear increasing trend of
Ly$\alpha$ flux with decreasing stellar $T_{\rm eff}$. M dwarfs produce, on
average, a nearly one order of magnitude higher Ly$\alpha$ flux in their HZ than 
do solar-like stars in their respective HZ.  Unfortunately, most of the stars used by 
{\em Linsky et al.} have no well-determined ages and thus a time-evolution determination 
is not possible. The Ly$\alpha$ flux evolution relationship of {\em Ribas et al.} 
(2005) for solar analogs is still up to date:
\begin{equation}
  F \approx 19.2 t_9^{-0.72}~{\rm [erg~s^{-1}~cm^{-2}]}
  \label{eq:Lya}
\end{equation}
where $F$ is the normalized flux at a distance of 1 AU and $t_9$ is the stellar
age in Gyr.

\bigskip
\noindent
\textbf{4.3 Evolution of High-Energy Radiation and Activity}
\bigskip

The strongest evolutionary change in magnetically induced stellar radiation
is seen in the high-energy range including EUV, X-ray and gamma-ray radiation.
The integrated luminosity in the soft X-ray (0.1--10~keV photon energy) and the
EUV (0.014--0.1~keV)
depends on the stellar rotation period, $P$, and therefore age; for a solar analog 
({\em G\"udel et al.}, 1997; {\em Sanz-Forcada et al.}, 2011),
\begin{eqnarray}
  L_\mathrm{X}   &\approx&  10^{31.05 \pm 0.12} P^{-2.64 \pm 0.12}~\mathrm{[erg\ s^{-1}]}\\
  L_\mathrm{X}   &\approx& (3\pm 1)\times 10^{28} t_9^{-1.5\pm 0.3}~\mathrm{[erg\ s^{-1}]}\\
  L_\mathrm{EUV} &\approx& (1.3\pm 0.3)\times 10^{29} t_9^{-1.24\pm 0.15}~\mathrm{[erg\ s^{-1}]}
\label{eq:activityrotationX}
\end{eqnarray}
($t_9$ is the stellar age in Gyr).
However, these dependencies only hold for sufficiently slow rotation. For young,
rapid rotators, magnetic activity and therefore the X-ray output reaches a 
saturation level. Empirically, $L_\mathrm{X} \approx
10^{-3}L_{\rm bol}$ for all late-type main-sequence and pre-main sequence stars.

These dependencies can be generalized to stars of other spectral class. For example,
using the Rossby number $Ro = P/\tau_c$ (where $\tau_c$ is the spectral-class and
age-dependent convective turnover time), one finds a unified description of the
$L_{\rm X} - Ro$ and $L_{\rm X} - P$ dependencies ({\em Pizzolato et al.}, 2003), 
namely $L_\mathrm{X} \approx 10^{-3}L_{\rm bol}$ for rapid rotators (young age) and 
$L_X \propto P^{-2}$ for slow rotators (more evolved stars).

\begin{figure}[tbp]
  \centerline{
      \includegraphics[scale=0.445]{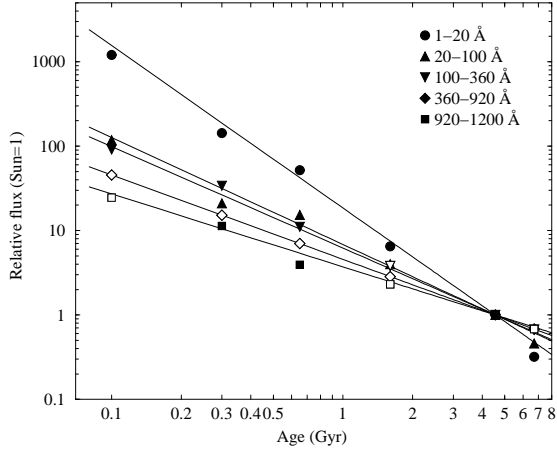}
  }
  \caption{\small Power-law decay of spectral output of a solar analog
  during its MS (age 0.05--10~Gyr) life,
  normalized to the present-day solar flux. The
  shortest-wavelength flux decays the fastest. 
  (From {\em Ribas et al.}, 2005; \copyright\ AAS. Reproduced with permission.)}
  \label{fig:decay}
\end{figure}

Combining these trends with the mass or spectral-type dependent spin-down rate of 
a star has important consequences for the stellar high-energy output in time.
Attempts to unify the X-ray decay law were described in {\em Scalo et al.} (2007),
{\em Guinan et al.} (2009) and {\em Mamajek \& Hillenbrand} (2008). M dwarfs, for example,
remain at the saturation level for much longer than solar-like
stars, up to order of a Gyr as for example seen in the Hyades cluster ({\em Stern et 
al.}, 1995), in contrast to a solar analog that leaves this regime
typically in the first 100~Myr.

A second evolutionary trend in X-rays is due to the average coronal temperature,
resulting in a long-term trend in spectral hardness. X-ray hardness is 
an important factor determining the penetration depth of X-rays into
a planetary atmosphere. While the present-day solar corona shows an average
temperature, $T_{\rm cor}$, of about 2~MK, young, magnetically active stars reveal
$T_{\rm cor}$ around 10~MK and even higher, implying a corresponding shift
of the dominant spectral radiation to harder photon energies ($\approx 1$~keV).
The empirical relation for solar-mass stars is ({\em Telleschi et al.}, 2005)
\begin{equation}
  L_\mathrm{X}  = 1.61\times 10^{26} T_\mathrm{cor}^{4.05\pm 0.25} \quad \mathrm{[erg\ s^{-1}]}.
\label{TLx}
\end{equation}
This relation holds similarly for pre-main sequence stars.

\begin{deluxetable}{lrrrr}
\tabletypesize{\small}
\tablecaption{Enhancement factors of X-ray/EUV/XUV/FUV fluxes in solar
  history$^a$ \label{table:enhancement}}
\tablewidth{0pt}
\tablehead{
Solar age  & Time before        & \multicolumn{3}{c}{\hrulefill\ Enhancement in\ \hrulefill} \\
(Gyr)      & present (Gyr)      & X-Rays (1\,--\,20\,\AA) & Soft-X (20\,--\,100\,\AA) &  FUV (920\,--\,1180\,\AA) \\
           &                    &                   & EUV (100\,--\,360\,\AA)         &  
	  } 
\startdata
0.1 	   & 4.5\phc		& 1600$^b$	  \phc    & 100\phantom{.5}\phc       &  25\phantom{.5}\phc \\
0.2 	   & 4.4\phc		&  400\phantom{$^b$}\phc  &  50\phantom{.5}\phc       &  14\phantom{.5}\phc \\
0.7 	   & 3.9\phc		&   40\phantom{$^b$}\phc  &  10\phantom{.5}\phc       &   5\phantom{.5}\phc \\
1.1 	   & 3.5\phc		&   15\phantom{$^b$}\phc  &   6\phantom{.5}\phc       &   3\phantom{.5}\phc \\
1.9 	   & 2.7\phc		&    5\phantom{$^b$}\phc  &   3\phantom{.5}\phc       &   2\phantom{.5}\phc \\
2.6 	   & 2.0\phc		&    3\phantom{$^b$}\phc  &   2\phantom{.5}\phc       &   1.6\phc \\
3.2 	   & 1.4\phc		&    2\phantom{$^b$}\phc  & 1.5 \phc                  &   1.4\phc \\
4.6 	   & 0\phantom{.5}\phc  &    1\phantom{$^b$}\phc  &   1\phantom{.5}\phc       &   1\phantom{.5}\phc \\
\enddata
\begin{list}{}{}
\item[$^{\mathrm{a}}$]{normalized to ZAMS age of 4.6\,Gyr before present}
\item[$^{\mathrm{b}}$]{large scatter possible due to unknown initial rotation period of Sun}
\end{list}
\vskip -1truecm
\end{deluxetable}

Short-wavelength (UV  to X-ray)
emission-line fluxes $F(T_{\rm max}, t)$ for maximum line formation temperatures 
$T_{\rm max}$ ($4 \la \log T_{\rm max} \la 7$) decay in time ($t_9$ in Gyr) as 
\begin{eqnarray}
  F(T_{\rm max}, t) &=& \alpha t_9^{-\beta} \\
  \beta &=& 0.32\log T_\mathrm{max} - 0.46
  \label{decaylaw}
\end{eqnarray}
({\em G\"udel}, 2007; similar relations hold for the continuum, where $T_{\rm max}$ corresponds to
$hc/(\lambda k)$, $h$ is the Planck constant, $k$ the Boltzmann constant, $c$ the 
speed of light, and $\lambda$ the continuum wavelength).
In other words, the decay is {\it steeper for shorter wavelengths,} implying the largest
enhancement factors for young stars for the shortest wavelengths (Fig.~\ref{fig:decay}). 
The average enhancement factors are summarized in Table~\ref{table:enhancement} for a solar 
analog throughout its main-sequence life. We emphasize that enhancement factors for individual
young stars may scatter around the given values because the rotation period may not have converged
to an age-independent value on the main sequence. We refer to the chapter of {\em Bouvier et al.}
in this volume for more details on the rotational evolution of stars.

\bigskip
\noindent
\textbf{4.4 The Evolution of Stellar Winds and Mass Loss}
\bigskip

In addition to being exposed to electromagnetic radiation from
their host stars, planets are also exposed to high-speed outflows
of particles from the stellar atmosphere.  For cool main sequence
stars like the Sun, stellar winds arise in the hot coronae that represent
the outermost atmospheres of the stars.  Although the mechanisms
of coronal heating and coronal wind acceleration remain hot topics
of research, {\em Parker} (1958) demonstrated long ago that once you
have a hot corona, a wind much like that of the Sun arises naturally
through thermal expansion.  Thus, any star known to have a
hot corona can be expected to possess a coronal wind something
like that of the Sun.  Observations from X-ray observatories such
as {\em Einstein}, {\em ROSAT}, {\em Chandra}, and {\em XMM-Newton}
have demonstrated that X-ray emitting coronae are ubiquitous among
cool main sequence stars, so coronal winds can be expected to be
ubiquitous as well.

Unfortunately, detecting and studying these winds is much harder
than detecting and studying the coronae in which they arise. Current
observational capabilities do not yet allow us to directly detect
solar-like coronal winds emanating from other stars. Bremsstrahlung radio 
emission may uncover ionized winds. An upper limit for the mass-loss rate of
$\dot{M}_\mathrm{w} \la 2\times 10^{-11}\,M_{\odot}~{\rm yr^{-1}}$ 
was inferred for the F5~IV-V subgiant Procyon ({\em Drake et al.}, 1993), and
$\dot{M}_\mathrm{w} \la 7\times 10^{-12}\,M_{\odot}~{\rm yr^{-1}}$ 
for the M dwarf Proxima Centauri (using a wind temperature $T_{\rm w} = 10^4$~K and a wind velocity
$v_{\rm w} = 300$~km~s$^{-1}$, {\em Lim et al.}, 1996). {\em Van den Oord and Doyle} (1997)
also found $\dot{M}_\mathrm{w} \la 10^{-12}\,M_{\odot}~{\rm yr^{-1}}$ for
observed dMe stars assuming $T_{\rm w} \approx 1~{\rm MK}$.
For young (few 100~Myr) solar analogs, {\em Gaidos et al.} (2000) found upper 
limits of $\dot{M}_\mathrm{w} \la (4-5)\times 10^{-11}\,M_{\odot}~
{\rm yr^{-1}}$. 

Sensitive upper limits can also be derived from the absence of radio attenuation
of coronal (flare or quiescent) radio emission by an overlying wind 
({\em Lim and White}, 1996). This method applies only to spherically symmetric 
winds. The most sensitive upper limits based on low-frequency radio
emission apply to the dMe star YZ CMi, with 
$\dot{M}_\mathrm{w} \la 5\times 10^{-14}\,M_{\odot}~{\rm yr^{-1}},
                    \la         10^{-12}\,M_{\odot}~{\rm yr^{-1}},\ {\rm and}\ 
                    \la         10^{-11}\,M_{\odot}~{\rm yr^{-1}}$
for $v_{\rm w} = 300$~km~s$^{-1}$ and $T_{\rm w} = 
10^4$~K, $10^6$~K, and $10^7$~K, respectively. 

Alternatively, neutrals of the interstellar medium flowing into an
astrosphere blown by a stellar wind may induce charge exchange that could
in principle be detected by the ensuing line emission at X-ray wavelengths.
Spatially resolved observations are needed, but again, no detection has been
reported yet (see {\em Wargelin and Drake}, 2001).

The so far only successful method relies not on detecting
the winds themselves, but their collision with the ISM ({\em Linsky and
Wood}, 1996).  The collision regions are called "astrospheres,"
analogous to the "heliosphere" that defines the solar wind's realm of
influence around the Sun.  Because the ISM surrounding the Sun is
partially neutral, the solar-wind/ISM collision yields populations of
hot hydrogen gas throughout the heliosphere created by charge exchange
processes.  The highest H\,{\sc i} densities are in the so-called ``hydrogen
wall'' just beyond the heliopause, hundreds of AU from the
Sun.  Ultraviolet spectra from the {\em Hubble Space Telescope} (HST)
have detected Lyman-$\alpha$ absorption from both the heliospheric
hydrogen wall ({\em Linsky and Wood}, 1996), and astrospheric hydrogen walls 
surrounding the observed stars ({\em Gayley et al.}, 1997;
{\em Wood et al.}, 2002, 2005).  In the case of astrospheres, the amount of absorption
is related to the strength of the stellar winds, so with guidance from
hydrodynamic models of the astrospheres, stellar mass loss rates have
been inferred from the astrospheric Lyman-$\alpha$ data. The number of 
astrospheric wind measurements is only 13, however, and many more 
measurements are required for further progress to be made in this area.

\begin{figure}[t!]
  \rotatebox{0}{\includegraphics[scale=0.83]{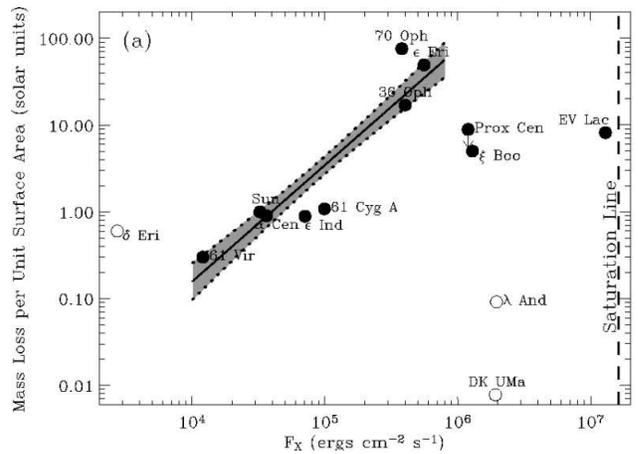}}
 \caption{\small Mass-loss rates per unit surface area vs.
  stellar X-ray surface fluxes. MS stars are shown by filled
  circles. (From {\em Wood et al.}, 2005; \copyright\ AAS. Reproduced with permission.)}  \label{fig:wind}
 \end{figure}

These observations imply that winds generally correlate with stellar activity 
({\em Wood et al.}, 2005; Fig.~\ref{fig:wind}). The mass-loss rate $\dot{\cal{M}}_\mathrm{w}$ 
per unit stellar surface correlates with the stellar X-ray surface flux $F_{\rm X}$;
correspondingly, the total mass-loss rate $\dot{M}_\mathrm{w}$ correlates 
with the X-ray luminosity $L_{\rm X}$ and therefore age, $t$,
\begin{eqnarray}
  \dot{M}_\mathrm{w} &\propto& L_\mathrm{X}^{1.34\pm 0.18}\\
  \dot{M}_\mathrm{w} &\propto& t^{-2.33\pm 0.55}
\end{eqnarray}
where for the second formula the activity-age relation has been used ({\em Wood et al.}, 2005). 
Extrapolating the above law up to the X-ray saturation limit at $F_\mathrm{X} \approx 2\times
10^7~{\rm erg\ cm^{-2}\ s^{-1}}$, we would infer 
$\dot{M}_\mathrm{w}$  of the youngest solar analogs to exceed the
present-day solar mass loss ($\dot{M}_{\odot} \approx 2\times 10^{-14}\,M_{\odot}~
{\rm yr^{-1}}$; e.g., {\em Feldman et al.}, 1977) by a factor of a thousand. 
However, there is some evidence that the most active stars with 
$F_\mathrm{X} \ga 8 \times 10^5~{\rm erg\ cm^{-2}\ s^{-1}}$ are
inconsistent with this correlation and may actually have surprisingly
weak winds ({\em Wood et al.}, 2005).  Mass loss appears to increase with 
activity up to about $\dot{M}_{\rm w}=100$ times the current solar wind, but
then appears to actually weaken towards even higher stellar
activity, down to about 10 times the current solar mass loss rate.
Possibly, the appearance of high-latitude
active regions (spots) with a  magnetic field more akin to 
a global dipole may inhibit wind escape ({\em Wood et al.}, 2005).

We emphasize that this method requires the presence of {\it neutrals}
(e.g., at least a partially neutral cloud) around the observed star.
If the interstellar medium around the star is fully ionized, no
hydrogen walls can form, and the absence of absorption features in Ly$\alpha$
will {\it not} indicate weak winds.

\bigskip
\noindent
\textbf{4.5 Stellar Flares}
\bigskip

A stellar flare is the result of short-term (minutes to hours) explosive energy
release tapped from non-potential energy in the coronal magnetic fields, ultimately
derived from convective energy at the stellar surface. During flares, plasma in
the stellar atmosphere is rapidly heated, and therefore 
fluxes increase across the electromagnetic spectrum, in very rare
cases exceeding the entire quiescent stellar output ({\em Osten et al.}, 2010).
Flare radiation amplitudes are particularly high in the short-wavelength
(XUV, gamma-ray) range, and also in the radio regime. 

Flares occur at rates depending on the flare amplitude or its released 
(radiative or total) energy, $E$, in the form of a power law, $dN/dE \propto E^{-\alpha}$,
where $\alpha \approx 2 \pm 0.4$ has been found by various investigators, i.e.,
small flares occur at a much higher rate than the rare big flares ({\em G\"udel}, 2004, and
references therein). 

Very energetic flares or ``superflares'' may contribute to the
ionization of upper planetary atmospheres (ionospheres) and upper-atmospheric
chemistry, thus potentially changing planetary habitable conditions. {\em Audard et al.}
(2000) found that the rate of large EUV flares exceeding some given radiative energy 
threshold is proportional to the average, long-term X-ray luminosity of the star,
suggesting a very high rate of superflares in young, active stars. This relation
at the same time indicates that superflares are not occurring at a higher rate
in M dwarfs than in solar analogs; it is the closer-in habitable zones around M dwarfs
combined with the slower decay of M-dwarf activity that subjects habitable planets
around M dwarfs to much higher flare-radiation doses for a longer time. 

{\em Schrijver et al.} (2012) investigated the presence of superflares on the Sun in historical
times, using natural archives (nitrate in polar ice cores and various cosmogenic radionuclids) 
together with flare statistics and the historical record of sunspots; they conclude that flares
in the past four centuries are unlikely to have exceeded the largest observed
solar flare with total energy of $\approx 10^{33}$~erg. 
The solar magnetic field (an similarly, stellar magnetic
fields) provides an upper bound to the maximum flare energy release simply by the
amount of free magnetic energy that could coherently be converted. This limit is
indeed about $\approx 10^{33}$~erg, a tenfold stronger flare requiring a spot coverage
of $10\%$, unobserved on the modern Sun ({\em Schrijver et al.}, 2012).

However, {\em Schaefer et al.} (2000) reported, from observations in various wavebands, nine 
``super-flares'' with total radiative energies of $10^{33}-10^{38}$~erg on solar (F-G type) analog 
stars, none of which is exceptionally young or extremely active, or a member of a close binary system.
Some of the flaring objects were even older than the Sun. They concluded that the average
recurrence time for superflares on such stars may be decades to centuries although
this would obviously be an overestimate for the Sun. With the new {\it Kepler} satellite data in 
hand, {\em Maehara et al.} (2012) found surprisingly many superflares with bolometric
energies of $10^{33}-10^{36}$~erg, with an estimated recurrence time of $\approx$800~yrs for
$10^{34}$~erg flares on old, solar-like stars. None of these stars is known to host
close-in exoplanets that could induce such energy release. 

The impact of stellar flares on the atmospheres of terrestrial extrasolar planets has been 
studied by, e.g., {\em Segura et al.} (2010) and {\em Grenfell et al.} (2012). The importance of 
such single events would be their potential to melt ice surfaces on habitable planets, the possible
temporary breakup of the ionosphere and ozone depletion after creation of nitrogen oxides in
the irradiated atmosphere; an event of $10^{36}$~erg (ionizing energy) was estimated to 
result in a loss of 80\% of the total ozone content for more than one year, with
the consequent increase in UV irradiation ({\em Schaefer et al.}, 2000).

\bigskip
\noindent
\textbf{4.6 Coronal Mass Ejections}
\bigskip

Apart from stellar winds, exoplanets may interact with coronal mass ejections (CMEs), occurring
sporadically and propagating within the stellar wind as large-scale plasma-magnetic structures.
The high speeds of CMEs (up to thousands of km/s), their intrinsic magnetic field and their increased density compared
to the stellar wind background make CMEs an active factor which strongly influences the planetary environments and
magnetospheres. Often, collisions of the close-orbit exoplanets with massive stellar CME
plasmas compress planetary magnetospheres much deeper towards the surface of the exoplanet. This would
result in much higher ion loss rates than expected during the usual stellar wind conditions (see Sect. 6 below).

 CMEs can be directly observed only on the Sun.
They are associated with flares and prominence eruptions and
their sources are usually located in active regions and prominence sites. The likelihood of CME events
increases with the size and power of the related flare event. Generally, it is expected that the frequent
and powerful flares on magnetically active flaring stars should be accompanied by an increased rate of
CME production. Based on  estimates of solar CME plasma density $n_{\rm CME}$,
using the in-situ spacecraft measurements (at distances $> 0.4$ AU) and the analysis of white-light
coronagraph images (at distances $\leq 30 R_{\rm Sun} \approx 0.14$ AU), {\em Khodachenko et al.} (2007a)
provided general power-law interpolations for the $n_{\rm CME}$ dependence on the orbital distance $d$ 
(in AU) to a star:
\begin{equation}
  n^{\rm min}_{\rm CME}(d) = 4.88 d^{-2.3},\quad \:
  n^{\rm max}_{\rm CME}(d) = 7.10 d^{-3.0}.
  \label{eq-nCME}
\end{equation}
Equations (\ref{eq-nCME}) identify a typical maximum-minimum range of $n_{\rm CME}$. The dependence of stellar
CME speed $v_{\rm CME}$ on $d$ can be approximated by the formula:
\begin{equation}
v_{\rm CME} = v_0 \left( 1 - {\rm exp}\left[{\frac{2.8 R_{\rm Sun} - d}{8.1 R_{\rm Sun}}}\right] \right)^{1/2} ,
  \label{eq-vCME}
\end{equation}
proposed in {\em Sheeley et al.} (1997) on the basis of tracking of several solar wind density enhancements
at close distances ($d < 0.1$ AU). For the approximation of average- and high- speed CMEs one may take in
Eq.~(\ref{eq-vCME}) $v_0 = 500$ km~s$^{-1}$ and $v_0 = 800$ km~s$^{-1}$, respectively. Besides of that, the average mass
of CMEs is estimated as $10^{15}$ g, whereas their average duration at distances of $\approx 0.05$ AU is close
to 8 hours.

Because of the relatively short range of propagation of most CMEs, they should strongly impact the 
planets in close orbits ($\lesssim 0.3$ AU). {\em Khodachenko et al.} (2007a) have found that for a critical 
CME production rate $f_{\rm CME}^{\rm cr} \approx 36$ CMEs per day (and higher), a close orbit exoplanet appears 
to be under continuous action of the stellar CMEs plasma, so that each next CME collides with the planet during the time 
when the previous CME is still passing over it, acting like an additional type of ``wind''. 
 Therefore, investigations of evolutionary paths of close-orbit 
exoplanets in potentially habitable zones around young active stars 
should take into account the effects of such relatively dense magnetic clouds (MCs) and CMEs, apart from the
ordinary winds.

% Sect. 5
\section{\textbf{ATMOSPHERES AND STELLAR RADIATION}}

The radiation environment during the active period of a young star is of great
relevance for the evolution and escape of planetary atmospheres. The visible radiation penetrates
through a planetary atmosphere down to its surface unless it is blocked
by clouds, aerosols or dust particles.
The average skin or effective temperature, $T_{\rm eff}$, of a planet can be estimated from the
energy balance between the optical/near-IR emission irradiating the
planet and the mid-IR thermal radiation that is lost to space:
\begin{equation}
T_{\rm eff}=\left[\frac{S(1-A)}{4\sigma}\right]^{0.25},
\end{equation}
with $S$ being the solar radiative energy flux at a particular orbit location of a planet, 
$A$ the bond albedo
describing the fraction of the radiation reflected from the planet,
and $\sigma$ the Stefan-Boltzmann constant. Most of the IR radiation emitted from Venus or Earth 
does not come from the surface; rather, it comes from the cloud top level on Venus or the mid-troposphere 
on Earth and yields $T_{\rm eff}\approx 220$~K for Venus and 255~K for Earth.
Above the mesopause, radiation with short wavelengths in the X-ray,
soft-X-ray and the extreme ultraviolet (XUV) part of the spectrum is absorbed in
the upper atmosphere, leading to dissociation of molecular
species, ionization and heating,
so that the thermosphere temperature $T_{\rm th} \gg T_{\rm eff}$.

\bigskip
\noindent
\textbf{5.1 Surface Climate and Chemical Processing}
\bigskip

The surface climate conditions on terrestrial planets are largely affected by the presence of greenhouse gases (main
contributors: H$_2$O, CO$_2$, O$_3$, CH$_4$, N$_2$O), absorbing and re-emitting infrared (IR) radiation and thereby heating
the lower atmosphere. The greenhouse effect on the surface of Earth is about 33~K, and about 460 K for the
dense CO$_2$ atmosphere of Venus. In the troposphere of the terrestrial planets, adiabatic cooling leads to decreasing
temperatures with height. In the case of Earth, the ozone (O$_3$) layer peaking at about 30 km height absorbs stellar
radiation around 250~nm, resulting in atmospheric heating and a stratospheric temperature inversion. This is not observed for
those terrestrial planets in our solar system which do not have sufficient amounts of ozone, i.e., Mars and Venus. In the
context of extrasolar planets, ozone is of special interest not only for its climatic impact, but also as a potential
biosignature which can be detected by spectroscopic absorption bands in the IR spectral range.

Ozone is formed in the Earth's stratosphere via photolytic processes from O$_2$ ({\em Chapman}, 1930)  operating at wavelengths
~$<$200 nm. The ozone molecule itself is also destroyed photolytically in the UVB and via HO$_x$ and NO$_x$ catalytic cycles
({\em Crutzen}, 1970), which are also favored by UV radiation. In the Earth's troposphere, O$_3$ is mainly produced by the smog
mechanism ({\em Haagen-Smit et al.}, 1952).

\begin{figure}[b!]
  \rotatebox{270}{\includegraphics[scale=0.37,clip=true]{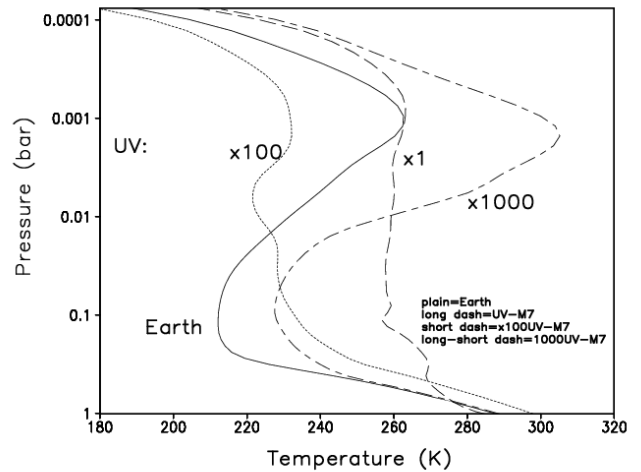}}
 \caption{\small Response of the atmospheric $T-p$ profile of an Earth-like planet in the HZ
 of an M7 red dwarf star to UV irradiation. The profiles are for 
 a star with very low UV flux (long dash, ``$\times 1$''), 
 100 and 1000 times this flux (dotted resp. long-short dash). The $T-p$ profile for the present
 Earth is also shown (solid). The cooling from $1\times$ to $100\times$ is due to the destruction 
 of CH$_4$ by UV, resulting in less heating. Very high UV levels lead to a temperature inversion in the ozone layer 
 (adapted from {\em Grenfell et al.}, 2014).  }  \label{fig:Tp}
 \end{figure}

Terrestrial exoplanets orbit host stars of different types, i.e., stars with different effective temperatures
and hence spectral energy distributions. Cool dwarf stars, such as M-types, exhibit less flux at optical and
near-UV wavelengths than the Sun down to about 200~nm, depending on stellar class. They can, however, be very
active in the UV range below 170-190~nm, hence at wavelengths relevant for photochemistry of O$_2$ and O$_3$ (Sect. 4.2).
Planets with oxygen-bearing atmospheres 
around M dwarf stars may therefore have weaker or stronger ozone
layers, depending on UV fluxes below about 200 nm (e.g., {\em Segura et al.}, 2005; {\em Grenfell et al.}, 2013). Cool
M dwarfs with reduced UV fluxes allow for less O$_2$ destruction and therefore less O$_3$ is produced by the
Chapman mechanism. In the case of very cool M dwarfs this production mechanism could be severely reduced and the
smog mechanism could take over even in the mid-atmosphere ({\em Grenfell et al.}, 2013). For Earth-like planets
around such cool and quiet M dwarfs atmospheric ozone abundances can be significantly reduced, whereas the
same kind of planet around an active M dwarf with high UV fluxes would show a prominent ozone layer ({\em Segura et al.}, 2005;
{\em Grenfell et al.}, 2014).  
The amount of stratospheric ozone on exoplanets around M dwarfs may therefore highly
depend on their UV activity.  Surface temperatures are somewhat higher on Earth-like planets (i.e., with modern Earth
type N$_2$-O$_2$ dominated atmospheres) orbiting M dwarfs for the same total stellar insolation, because,
e.g., Rayleigh scattering contributes less to the planetary albedo owing to the $\lambda^{-4}$-dependence of
the Rayleigh scattering coefficient. The largest impact of the different stellar energy flux distribution of
cool stars on a planet's $T-p$ profile is, however, seen in the mid-atmosphere. Earth-like terrestrial
exoplanets around M dwarfs do not show a strong stratospheric temperature inversion, even if a thick ozone
layer is present in the mid-atmosphere ({\em Segura et al. 2005}; {\em Rauer et al.}, 2011; {\em Grenfell et al.}, 2013). For
example, as discussed above, some active M dwarfs with very high fluxes below 200~nm can efficiently destroy
oxygen to form a strong ozone layer ({\em Segura et al.}, 2005), but still show no stratospheric $T$-inversion (Fig.~\ref{fig:Tp}). 
This is primarily caused by the reduced near-UV flux of M dwarf stars around 250 nm which is relevant for
stratospheric heating. 
In Fig.~\ref{fig:Tp},  increasing UV (from an inactive M7 dwarf - ``$\times 1$'' - to 1000$\times$ stronger) leads to middle atmosphere 
cooling because UV stimulates OH which reduces CH$_4$ (an important heater). On increasing UV further, however, 
O$_3$ is stimulated which leads to strong stratospheric heating. Results suggest that a strong $T$-inversion is 
less relevant for surface climate but can make the detection of atmospheric absorption bands more difficult 
(e.g., {\em Rauer et al.}, 2011; {\em Kaltenegger et al.}, 2011; {\em Grenfell et al.}, 2014).
In summary, UV can, on the one hand, enhance O$_3$, 
hence heat the middle atmosphere, but on the other hand it can stimulate OH, hence 
remove CH$_4$, which has the opposite effect - which effect is strongest depends 
on the amount of O$_3$, CH$_4$ and the UV intensity.

Other atmospheric species relevant for climate, habitability and biosignatures, such as water, methane, or
nitrous oxide are also affected by the different host star spectral energy distributions. Water and methane are
strong greenhouse gases and therefore relevant for habitability. Water is produced in the stratosphere via
methane oxidation (CH$_4$ + 2O$_2$ $\rightarrow$ CO$_2$ + 2H$_2$O) and destroyed by photolysis. Earth-like
planets around M dwarf stars show somewhat higher atmospheric water abundances  for the same
incident stellar flux due to higher surface
temperatures and reduced stratospheric photolytic destruction. In addition, the production of O$^1$D is
reduced due to the reduced M dwarf UV flux, affecting the production of OH via: H$_2$O + O$^1$D $\rightarrow$
2OH. This in turn leads to reduced methane destruction, hence increased methane abundances in the
atmosphere (e.g., {\em Segura et al.}, 2005; {\em Rauer et al.}, 2011). Also the good
biosignature  N$_2$O (good in the sense of attributability of spectral features to biogenic activity) 
could be significantly increased for planets around cool M dwarfs ({\em Rauer et al.}, 2011; 
{\em Seager et al.}, 2013).
This is interesting because N$_2$O is normally not detectable for Earth-like planets around solar-like stars
due to its weak spectral features, but may be detectable for planets orbiting M dwarf stars. These examples
illustrate how the central star spectral energy distribution affects atmospheric chemistry, abundances and
the strength of potentially observable atmospheric molecular absorption bands. The discussion has just
started, and many more parameter studies are needed to identify the most dominant effects on observable
signals.

Clearly, the visibility of indicators for habitability (e.g., H$_2$O, CO$_2$, CH$_4$) in spectra obtained 
from terrestrial exoplanets are significantly affected by the central star spectral type and activity 
(e.g., {\em Segura et al.}, 2005; {\em Selsis et al.}, 2008; {\em Segura et al.}, 2010;
{\em Rauer et al.}, 2011; {\em Hedelt et al.}, 2013; {\em Grenfell et al.}, 2014; {\em Rugheimer et al., 
2013}). This holds in particular for M dwarfs which are
favorable targets for transit spectroscopy during primary and secondary eclipse because of their improved
star/planet contrast. It has become evident that a good characterization of the host star stellar energy
distribution and its temporal variability is crucial for the interpretation of absorption spectra and
identification of potential false-positive signals.

{\it Clouds} are a common phenomenon in the atmospheres of many terrestrial bodies of our Solar System  (e.g., Venus,
Earth, Mars, Titan) and are also expected in atmospheres of terrestrial extrasolar planets. They play a major
role in the climate of terrestrial planets by affecting the atmospheric energy budget in several competing
ways. Clouds can scatter a fraction of the incident stellar radiation back to space, resulting in
atmospheric cooling. On the other hand, by trapping thermal radiation within the lower atmosphere clouds can
yield a greenhouse effect. This greenhouse effect can occur either by absorption and re-emission at their
local temperature (classical greenhouse effect) or by scattering thermal radiation back toward the planetary
surface (scattering greenhouse effect). In the Earth's atmosphere, for example, low-level water clouds usually
cool the surface whereas high-level water ice clouds can lead to a heating effect. The net climatic effect of
a cloud is largely determined by the wavelength-dependent optical properties of the cloud particles (e.g.,
absorption and scattering cross-sections, single-scattering albedo, asymmetry parameter, or scattering phase
function). These properties can differ considerably for different cloud-forming species, particle sizes, and
shapes (see {\em Marley et al.}, 2013, for a broad review on clouds in the atmospheres of extrasolar planets).

The climatic effect of clouds has been discussed, for example, as possible solutions to the Faint Young Sun
paradox (e.g., {\em Rondanelli and Lindzen}, 2010;  {\em Goldblatt and Zahnle}, 2011a; see Sect. 7.1 below), for planets at the inner boundary of
the habitable zone (e.g., {\em Kasting}, 1988), and their effect on exoplanet spectroscopy of planets at different
ages (e.g., {\em Kaltenegger et al.}, 2007). The climatic effect of clouds in atmospheres of Earth-like exoplanets
and the resulting effects on the potential masking of molecular absorption bands in the planetary spectra
has been studied again recently by, e.g., {\em Kitzmann et al.} (2010, 2011a, 2011b, 2013), {\em Zsom et al.} (2012), {\em Rugheimer
et al.} (2013), and {\em Vasquez et al.} (2013a, 2013b).

The impact of clouds on the climate of particular exoplanets is, however, difficult to evaluate without
observational constraints on their cloud cover or all the microphysical details (e.g., condensation nuclei)
which determine the cloud particle sizes and shapes ({\em Marley et al.}, 2013).

Active M dwarfs or even young solar-like stars may exhibit strong stellar cosmic-ray emissions. So-called air
shower events lead to secondary electron emissions which can break up strong atmospheric molecules like
N$_2$. This can perturb the atmospheric photochemistry, hence affect biosignature molecules and their
detectability (see, e.g., {\em Segura et al.} (2010) and {\em Grenfell et al.} (2007, 2012) for a discussion on the effect on
ozone and other spectral biosignatures). Results suggest for the extreme "flaring case" (see also Sect. 4.5)  that ozone may be
strongly removed by cosmic ray induced NOx ({\em Grenfell et al.}, 2012) whereas the nitrous oxide biosignature may
survive. Again, this illustrates the importance of characterizing stellar activity when discussing atmospheric
molecular signatures.

\bigskip
\noindent
\textbf{5.2 Upper Atmosphere XUV Heating and IR-Cooling}
\bigskip

Planets irradiated by high XUV radiation experience heating of the upper atmosphere,
related expansion and thermal escape. Ionizing XUV radiation at short wavelengths ($\lambda \leq$ 102.7~nm) and dissociating radiation (UV) as well 
as chemical heating in exothermic 3-body reactions are responsible for the heating of the upper planetary atmospheres (e.g., {\it Lammer}, 
2013). The volume heat production rate $q$ caused by the absorption of
the stellar XUV radiation can then be written as (e.g., {\em Erkaev et al.}, 2013; {\em Lammer et al.}, 2013b)
\begin{equation}
q_{\rm XUV} = \eta n \sigma_{\rm a}J e^{-\tau},
\end{equation}
with the energy flux related to the corresponding wavelength range outside
the atmosphere, $J$, the absorption cross-section of atmospheric species, $\sigma_{\rm a}$, the optical depth, $\tau$,
and the number density, $n$. In today's Venus' or Earth's atmosphere, thermospheric heating processes occur near the homopause in the lower thermosphere,
where the eddy diffusion coefficient is equal to the molecular diffusion coefficient, at an altitude of $\sim110$ km (e.g., {\it von Zahn et al.}, 1980).
Depending on the availability of the main atmospheric constituents and minor species, the thermospheric temperature can be reduced 
by IR emission in the vibrational-rotational bands of molecules such as CO$_2$, O$_3$, H$_3^+$, etc.

The fraction of the absorbed stellar XUV radiation transformed into thermal energy is the so-called heating efficiency $\eta$;
it lies  in the range of $\sim$15--60\% ({\em
Chassefi\`{e}re}, 1996a; {\em Yelle}, 2004; {\em Lammer et al.}, 2009; {\em Koskinen et al.}, 2013). 
Recent calculations by {\em Koskinen
et al.} (2013) estimate the total heating rate based on the absorption of stellar XUV
radiation, photoelectrons and photochemistry for the hydrogen-rich ``hot Jupiter''
HD 209458b with an orbit of 0.047~AU around a solar like star; they  found that the heating efficiency for
XUV fluxes $\sim$450 times today's solar flux at 1~AU is
most likely between $\sim$40--60\%. For hydrogen-rich exoplanets exposed to such high
XUV fluxes most molecules such as H$_2$ in the thermosphere are dissociated; thus, IR cooling
via H$_3^+$  produced by photochemical reactions with H$_2$ molecules becomes less important
({\em Koskinen et al.}, 2007). However, depending on the availability of  IR-cooling molecules
in planetary atmospheres at larger orbital distances exposed to lower XUV fluxes,
the heating efficiency may be closer to the lower value of $\sim$15\%.

\bigskip
\noindent
\textbf{5.3 XUV Induced Thermal Atmospheric Escape}
\bigskip

{\em Tian et al.} (2008) studied the response of Earth's upper atmosphere
to high solar XUV fluxes in more detail and discovered that
one can classify the thermosphere response into two thermal
regimes.
In the first regime, the thermosphere is in hydrostatic equilibrium. In
this stage the bulk atmosphere below the exobase can
be considered as static similar to the thermospheres
of the solar system planets today. However, if the
star's EUV flux was/is much higher
than that of the present Sun, or if the planet's atmosphere is hydrogen-rich ({\em Watson et al.}, 1981),
the thermosphere can enter a second regime.
There, the upper atmosphere is heated to such temperatures
that the thermosphere becomes non-hydrostatic.
In the hydrodynamic flow regime, the major
gases in the thermosphere can expand very efficiently
to several planetary radii above the planetary surface. As
a result light atoms experience high thermal escape rates
and expand to large distances.

In the hydrostatic regime the upper atmosphere experiences classical
Jeans escape where
particles populating the high-energy tail of a Maxwell distribution
at the exobase level have escape energy so that they are lost from the planet.
The so-called Jeans parameter, $\lambda_{\rm J}(r)$, can generally be expressed as (e.g., {\em Chamberlain}, 1963)
\begin{equation}
\lambda_{\rm J}(r) =\frac{GM_{\rm pl}m}{rkT(r)},
\end{equation}
with the gravitational constant $G$, the planetary mass $M_{\rm pl}$, the mass of an atmospheric species
$m$, the Boltzmann constant $k$, and the upper atmosphere temperature $T$
as a function of planetary distance $r$. As long as $\lambda(r)$ is $>30$, an atmosphere is
bound to the planet. For lower values, thermal escape occurs and can become very high
if $\lambda(r) \sim$2--3.5 ({\em Volkov and Johnson}, 2013), depending on the atmospheric main species,
the XUV flux, and planetary parameters. 

In the non-hydrostatic regime, still not all atoms may
reach escape velocity at the exobase level. In such cases one can expect
hydrodynamically expanding thermospheres where the loss of the upward flowing atmosphere
results in a strong Jeans-type or controlled hydrodynamic escape but not in hydrodynamic blow-off
where no control mechanism influences the escaping gas.

Only for $\lambda(r)< 1.5$ does hydrodynamic blow-off occur, and the escape becomes uncontrolled.
This happens when the mean thermal energy of the gases at the exobase level
exceeds their gravitational energy. Hydrodynamic blow-off can be
considered as the most efficient atmospheric escape process. In this extreme condition,
the  escape is very high because the whole exosphere evaporates and
will be refilled by the upward flowing planetary gas of the dynamically expanding
thermosphere as long as the thermosphere can remain in this extreme condition.
In such cases the thermosphere starts, accompanied by adiabatic cooling, to dynamically expand to several planetary radii
({\em Sekiya et al.}, 1980; {\it Watson et al.}, 1981; {\em Chassefi\`{e}re}, 1996a; {\em Tian et al.}, 2005, 2008; {\em Kulikov et al.}, 2007).

To study the atmospheric structure of an upward flowing, hydrodynamically expanding
thermosphere, one has to solve the set of the hydrodynamic equations.
For exoplanets orbiting very close to their host stars,
one cannot neglect external forces such as gravitational effects related to the
Roche lobe ({\em Penz et. al.}, 2008).

\bigskip
\noindent
\textbf{5.4 XUV Powered Escape of H-Rich Protoatmospheres}
\bigskip

The most efficient atmospheric escape period starts after the planetary 
nebula dissipated and the protoplanet with its nebula-captured hydrogen-dominated envelope is exposed to the
extreme radiation and plasma environment of the young star. The time period of this extreme escape process depends generally
on the host star's evolving XUV flux, the planet's orbit location, its gravity, and the main atmospheric
constituents in the upper atmosphere. Some planets can be in danger of being stripped of their whole atmospheres;
on the other hand, if the atmospheric escape processes are too weak, a planet may have problems
getting rid of its protoatmosphere ({\em Lammer et al.}, 2011b, 2011c; {\em Lammer}, 2013; {\em Erkaev et al.,} 2013; {\em Kislyakova et al.}, 2013).
Both scenarios will have essential implications for habitability.

That the early atmospheres of terrestrial planets were even hydrogen-dominated or contained more
hydrogen than nowadays was considered decades ago by researchers such as {\em Holland} (1962),
{\em Walker} (1977), {\em Ringwood} (1979), {\em Sekiya et al.} (1980, 1981), {\em Watson et al.}
(1981), and more recently by {\em Ikoma and Genda} (2006). The capture or accumulation of hydrogen envelopes depends on the planetary
formation time, the nebula dissipation time, the depletion factor of dust grains in the nebula, the nebula opacity,
orbital parameters of other planets in a system, the gravity of the protoplanet, its orbital
location, as well as the host star's radiation and plasma environment (e.g., {\em Ikoma and Genda}, 2006;
{\em Rafikov}, 2006). Depending on all of these parameters,  theoretical studies let us expect that
fast growing terrestrial planets from Earth to ``super-Earth''-type may capture from tens to hundreds or even several
thousands of Earth ocean equivalent amounts of hydrogen around their rocky cores (e.g., {\em Hayashi et al.},
1979; {\em Mizuno}, 1980; {\em Wuchterl}, 1993; {\em Ikoma et al.}, 2000; {\em Ikoma and Genda}, 2006;
{\em Rafikov}, 2006),  making them really ``mini-Neptunes''. The recent discoveries of ``mini-Neptunes''
with known sizes and masses indicate that many of them have rocky cores surrounded by a significant
amount of hydrogen envelopes (e.g., {\em Lissauer et al.}, 2011; {\em Erkaev et al.}, 2013, 2014; {\em Kislyakova et al.},
2013; {\em Lammer}, 2013).

{\em Adams et al.} (2008), {\em Mordasini et al.} (2012), and {\em Kipping et al.} (2013) studied the mass-radius 
relationship of planets with primordial hydrogen envelope. The H envelope mass fraction can be
written as
\begin{equation}
f=\frac{M_{\rm atm}}{M_{\rm atm} + M_{\rm core}},
\end{equation}
with atmosphere mass $M_{\rm atm}$ and core mass $M_{\rm core}$. Their results indicate that for a given
mass, there is a considerable diversity of radii, mainly due to different bulk compositions, reflecting different
formation histories. According to {\em Mordasini et al.} (2012)
a protoplanet inside the habitable zone at 1 AU,
 with $T_{\rm eff}=250$~K, with a core mass of 1$M_{\rm \oplus}$ and a hydrogen envelope mass function $f$ of
 $\sim0.01$ could have a radius of about $\sim2R_\oplus$ depending on its age, orbital distance
and activity of its host star.

Additional to such nebula-based hydrogen envelopes, catastrophically outgassed volatile-rich steam atmospheres
which depend on the impact history and the initial volatile content of a planet's interior could also be formed
after a planet finished its accretion (e.g., {\em Elkins-Tanton and Seager}, 2008; {\em Elkins-Tanton}, 2011).
{\em Zahnle et al.} (1988) studied the evolution of steam atmospheres as expected on the early Earth and found for
surface temperatures close to 2000 K atmospheric temperatures of the order of $\sim$1000 K.
Recently, {\em Lebrun et al.} (2013) studied the lifetime of catastrophically outgassed steam atmospheres of early Mars,
Earth and Venus and found that water vapor condenses to an ocean after 0.1, 1.5, and 10 Myr, respectively. {\em Hamano et al.} (2013)
defined two planet categories:
a type I planet formed beyond a certain critical distance from the host star, solidifying, in agreement with
{\em Lebrun et al.} (2013), within several million years;
and type II planets formed inside the critical distance where a
magma ocean can be sustained for longer time periods of up to 100 Myr, even with a larger initial amount of water.

\begin{figure}[t]
\begin{center}
\includegraphics[width=8.3cm]{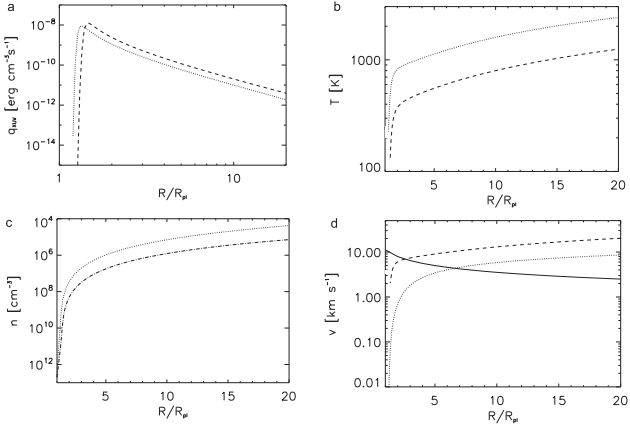}
\caption{\small Profiles of the volume heating rate (a), thermosphere temperature (b), atmospheric density (c) and outflow velocity (d)
as a function of altitude, for an atomic-hydrogen dominated upper atmosphere of an Earth-like planet with a heating efficiency of 15\%,  
subject to an XUV exposure 100 times the present Sun's at 1 AU.  The dotted and dashed lines correspond to a
temperature $T_{\rm eff}$ of 250 K at the base of the thermosphere with a lower thermosphere near the planetary surface (dotted line) and one
for a hydrogen-dominated protoatmosphere with an envelope mass fraction $f$ of 0.01 and a corresponding radius of $2R_{\oplus}$
({\em Mordasini et al.}, 2012). The solid line in (d) corresponds to the escape velocity as a function of distance.}
\label{escape}
\end{center}
\end{figure}

Fig.~\ref{escape} compares the hydrodynamically expanding hydro\-gen-dominated 
thermosphere structure for two cases: an Earth-like planet with $T_{\rm eff}$ of 250~K,
located within a G-star habitable zone at 1 AU, exposed to an XUV flux 100 times the
present-day Sun's on the one hand, and a similar planet (dashed line) 
with a more extended nebular-based hydrogen envelope mass fraction of $\sim$0.01 with a 
corresponding radius of $\sim2R_\oplus$ ({\em Mordasini et al.},  2012), on the other hand. 

For both scenarios shown in Fig. ~\ref{escape}  the transonic points are reached below the 
exobase levels so that the considered atmospheres are in the controlled hydrodynamic regime.

The thermal escape rate with a heating efficiency of 15\% for the case where the base of the thermosphere is close to the
planetary radius is $\sim 3.3\times 10^{31}$ s$^{-1}$; for the hydrogen-dominated protoatmosphere with $f=0.01$ and the
corresponding radius at $\sim2R_\oplus$, it is $\sim 5.0\times 10^{32}$ s$^{-1}$. The total losses of hydrogen in equivalent
amounts of Earth ocean contents (1 EO$_{\rm H} \sim 1.5\times 10^{23}$ g)
for the first 100 Myr would be $\sim 1 $EO$_{\rm H}$ for the case with the smaller lower thermosphere distance and
$\sim 15.5$EO$_{\rm H}$ for the extended hydrogen dominated protoatmosphere.

One can see from these results that an equivalent amount of only a few EO$_{\rm H}$ may be lost during the first 100 Myr (when the stellar XUV radiation
is saturated at high levels); one may expect that terrestrial planets growing rapidly to
Earth-mass or ``super-Earth-mass'' bodies inside the nebula  may have a problem with losing their captured hydrogen envelopes.
Some planets may keep tens, hundreds or even thousands of EO$_{\rm H}$ ({\em Lammer}, 2013; {\em Lammer et al.}, 2013b).

It is obvious that early Venus or Earth did not capture such huge amounts of nebula gas and
most likely did not accrete so fast by collisions with planetary embryos after the nebula disappeared ({\em Albar\`{e}de and Blichert-Toft, 2007}).

For all atmospheric escape processes the age of the host star and planet
as well as the host star's activity are very important. After the planet's origin, the briefly discussed hydrodynamic expansion and
outflow due to the extreme stellar and protoplanetary conditions result in the most efficient atmospheric
escape process. During this phase no intrinsic planetary magnetic field will protect the expanded
upper atmosphere from ion erosion by the stellar wind. The hydrodynamic outflow changes with decreasing XUV
flux and cooling of the lower thermosphere from the hydrodynamic to the hydrostatic
regime. For this later phase, a recent study by {\em Kislyakova et al.} (2013) indicates that stellar wind induced ion pick-up of
light species such as atomic hydrogen is always less efficient than thermal escape. However, after the XUV flux
of a young and active host star decreases to more moderate XUV flux levels  $<5$ times that of the
present solar value, non-thermal escape processes such as ion pick-up, detached plasma clouds, sputtering, and
photochemical losses especially for planets with low gravity become relevant for
heavier atmospheric species such as oxygen, carbon and nitrogen.

% Sect. 6
\section{\textbf{MAGNETOSPHERIC PROTECTION}}

As a consequence of upper-atmospheric heating by soft X-rays and EUV radiation, the expanding upper atmospheres may reach
and even exceed the boundaries of possible planetary magnetospheres. In this case the upper atmosphere will be directly exposed to the plasma
flows of the stellar wind and coronal mass ejections (CMEs) with the consequent loss due to ion pick-up, as well as sputtering, and different
kinds of photo-chemical energizing mechanisms, all contributing to {\em non-thermal} atmospheric mass-loss process
({\em Lundin et al.}, 2007 and references therein; {\em Lichtenegger et al.}, 2009; {\em Lammer}, 2013, and references therein).
As a crucial parameter appears here the size of the planetary magnetosphere. Altogether, this makes the
planetary magnetic field, as well as the parameters of the stellar wind (mainly its density $n_{\rm w}$ and speed $v_{\rm w}$) very
important for the processes of non-thermal atmospheric erosion and mass-loss of a planet, eventually affecting the whole evolution of its
environment and possible habitability.

\bigskip
\noindent
\textbf{6.1 The Challenge of Magnetospheric Protection}
\bigskip

The magnetosphere acts as an obstacle that interacts with the stellar wind, deflecting it and protecting planetary
ionospheres and upper atmospheres against the direct impact of stellar wind plasmas and energetic particles (e.g., cosmic rays).
For an efficient magnetospheric protection of a planet, the size of its magnetosphere characterized by the magnetopause stand-off distance
$R_{\rm s}$ should be much larger than the height of the outer layers of the exosphere. The value of $R_{\rm s}$ is determined by the balance between the
stellar wind ram pressure and the planetary magnetic field pressure at the substellar point ({\em Grie{\ss}meier et al.}, 2004;
{\em Khodachenko et al.}, 2007a). In most studies related to exoplanetary magnetospheric protection,
highly simplified planetary dipole-dominated magnetospheres have been assumed. This means that only the intrinsic magnetic dipole
moment of an exoplanet, $\cal{M}$, and the corresponding magnetopause electric currents (i.e., the "screened magnetic dipole" case) are considered
as the major magnetosphere forming factors. In this case, i.e., assuming $B(r) \propto \mathcal{M} / r^3$, the value of $R_{\rm s}$ has been
defined by the following expression ({\em Baumjohann and Treumann}, 1997),
\begin{equation}
  R_{\rm s} \equiv R_{\rm s}^{(dip)} = \left[ \frac{\mu_{\rm 0} f_{\rm 0}^2 {\cal{M}}^2}{8 \pi^2 \rho_{\rm w} \tilde{v}_{\rm w}^2} \right]^{1/6},
   \label{eq-1}
\end{equation}
where $\mu_{\rm 0}$ is the diamagnetic permeability of free space, $f_{\rm 0} \approx 1.22$ is a form-factor of the magnetosphere caused by
the account of the magnetopause electric currents, $\rho_{\rm w} = n_{\rm w} m$ is the mass density of the stellar wind, and 
$\tilde{v}_{\rm w}$ is the relative velocity of the stellar wind plasma which includes also the planetary orbital velocity.

The planetary magnetic field is generated
by the magnetic dynamo. The existence and efficiency of the dynamo are closely related to the type of the planet and its interior structure.
Limitations of the existing observational techniques make direct measurements of the magnetic fields of exoplanets impossible. Therefore, an estimate
of $\mathcal{M}$ from simple scaling laws found from an assessment of different contributions in the governing equations of planetary magnetic dynamo theory
({\em Farrell et al.}, 1999; {\em S\'anchez-Lavega}, 2004; {\em Grie{\ss}meier et al.}, 2004; {\em Christensen} 2010) is widely used. Most of these scaling
laws reveal a connection between the intrinsic magnetic field and rotation of a planet. More recently, {\em Reiners and Christensen} (2010), based on
scaling properties of convection-driven dynamos ({\em Christensen and Aubert}, 2006), calculated the evolution of average magnetic fields of
``Hot Jupiters'' and found that (a) extrasolar gas giants may start their evolution with rather high intrinsic magnetic fields, which then decrease
during the planet lifetime, and (b) the planetary magnetic moment may be independent of planetary rotation ({\em Reiners and Christensen}, 2010).
In the case of \emph{rotation-dependent} dynamo models, the estimations of $\mathcal{M}$ give rather small values for tidally locked close-orbit
exoplanets, resulting in small sizes of dipole-dominated magnetospheres, $R_{\rm s} = R_{\rm s}^{(dip)}$, compressed by the stellar wind plasma
flow. The survival of planets in the extreme conditions close to active stars  is often used as an
argument in favor of \emph{rotation-independent} models, or in favor of a generalization of the dipole-dominated magnetosphere
model.

{\em Khodachenko et al.} (2007b) studied the mass loss of the Hot Jupiter HD~209458b due to the ion pick-up mechanism caused by stellar CMEs
colliding with the planet. In spite of the sporadic character of the CME-planetary collisions for the moderately active host star
it has been shown that the integral action of the stellar CME impacts over the exoplanet's lifetime can produce a significant effect
on the planetary mass loss. The estimates of the non-thermal mass loss of a weakly magnetically protected  HD~209458b  due
to stellar wind ion pick-up suggest significant and sometimes unrealistic values -- losses up to several tens of planetary masses $M_{\rm p}$ during
a planet's life time ({\em Khodachenko et al.}, 2007b). Because multiple close-in giant exoplanets exist, comparable in mass and size
to Jupiter, and because it is unlikely that all of them began their life as much more ($>$ ten times) massive objects, one
may conclude that additional factors and processes must be considered to avoid full planetary destruction.

\bigskip
\noindent
\textbf{6.2 Magnetodisk-Dominated Magnetospheres}
\bigskip

The magnetosphere of a close-orbit exoplanet is a complex object whose formation depends on different external and internal factors. These factors
may be subdivided into two basic groups: (a) \emph{stellar factors}, e.g., stellar radiation, stellar wind plasma flow, stellar magnetic field and
(b) \emph{planetary factors}, e.g., the type of planet, its orbital characteristics, its escaping material flow, and its magnetic field. The structure
of an exoplanetary magnetosphere depends also on the speed regime of the stellar wind plasma relative the planet ({\em Erkaev et al.}, 2005;
{\em Ip et al.}, 2004). In particular, for an exoplanet at sufficiently large orbital distance where the stellar wind is super-sonic and
super-Alfv\'{e}nic, i.e., where the ram pressure of the stellar wind dominates the magnetic pressure, a Jupiter-type magnetosphere with a bow
shock, magnetopause, and magnetotail is formed. At the same time, in the case of an extremely close orbital location of an exoplanet (e.g.,
$d < 0.03$~AU for a solar analog), where the stellar wind is still being accelerated and remains sub-magnetosonic and
sub-Alfv\'{e}nic ({\em Ip et al.}, 2004; {\em Preusse et al.}, 2005), an Alfv\'{e}nic wing-type magnetosphere without a shock in the upstream
region is formed.  The character of the stellar wind impact on the immediate planetary plasma environment and the atmosphere is different for
the super- and sub-Alfv\'{e}nic types of the magnetosphere. Here, we briefly discuss moderately short-orbit giant planets
around solar-type stars subject to a super-Alfv\'{e}nic stellar wind flow, i.e., the magnetospheres having in general
a bow shock, a magnetopause, and a magnetotail, similar to Jupiter. Equivalent configurations are expected for Earth-like planets with
evaporating atmospheres.

To explain the obvious survival and sufficient magnetospheric protection of close-orbit Hot Jupiters under the extreme conditions of their
host stars, {\em Khodachenko et al.} (2012) proposed a more generic view of a Hot Jupiter magnetosphere. A key element in the proposed
approach consists of taking into account the upper atmosphere of a planet as an expanding dynamical gas layer heated and ionized by
the stellar XUV radiation ({\em Johansson et al.}, 2009; {\em Koskinen et al.}, 2010). Interaction of the outflowing plasma with the
rotating planetary magnetic dipole field leads to the development of a current-carrying magnetodisk surrounding the exoplanet. The inner edge
of magnetodisk is located at the so called Alfv\'{e}nic surface ($r=R_{\rm A}$) where the kinetic energy density of the moving plasma
becomes equal to the energy density of the planetary magnetic field ({\em Havnes and Goertz}, 1984).
%; {\em Andre et al.}, 1988).

%-------------------------------------------------------------------------------Fig.4.1
\begin{figure}[t!]
\begin{center}
\includegraphics[width=0.4\textwidth, clip=]{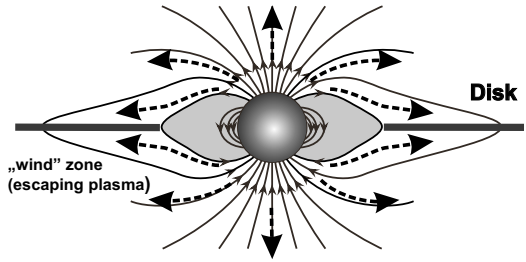}
\caption{\small Schematic view of magnetodisk formation: The expanding plasma flows 
along the magnetic field lines and changes the initial dipole topology 
of the field by the creation of a current disk.
In the dead zone (shadowed region), plasma is locked by the strong 
magnetic field;  in the wind zone, plasma escapes along the open field 
lines in the form of an expanding wind.  }
\label{fig:Fig4.1}
\end{center}
\end{figure}
%-------------------------------------------------------------------------------

Two major regions, separated by the Alfv\'{e}nic surface at $r = R_{\rm A}$, with different topologies of the magnetic field may be distinguished
in the magnetosphere of a Hot Jupiter driven by the escaping plasma flow ({\em Mestel}, 1968; {\em Havnes and Goertz}, 1984).
The first region corresponds to the inner magnetosphere, or so-called ``dead zone'', filled
with closed dipole-type magnetic field lines. The magnetic field in the ``dead zone'' is strong enough to keep plasma locked with the
planet. In the second region, so-called ``wind zone'', the expanding plasma drags and opens the magnetic field lines (see
Fig.~\ref{fig:Fig4.1}). The plasma escaping along field lines beyond the Alfv\'{e}nic surface not only deforms and stretches the
original planetary dipole field, but also creates a thin disk-type current sheet in the equatorial region. Altogether, this leads to the
development of a new type of magnetodisk-dominated magnetosphere of a Hot Jupiter, which has no analog among the solar system
planets ({\it Khodachenko et al.}, 2012). It enlarges the magnetospheric obstacle, thus {\it helping to suppress non-thermal escape 
processes.}

\bigskip
\noindent
\textbf{6.3 Wind-Exosphere-Magnetosphere Interaction}
\bigskip

As discussed before, extended hydrogen exospheres will be produced if hydrogen is the dominating constituent
in the upper atmosphere. {\em Vidal-Madjar et al.} (2003) were the first to observe the transiting exoplanet HD 209458b
with the HST/STIS-instrument and discovered a 15$\pm$4\% intensity drop in the stellar Lyman-$\alpha$ line in the high
velocity part of the spectra. The estimate of the absorption rate has later been carried out independently by
{\em Ben-Jaffel} (2007) who found a lower absorption rate of about 8.9$\pm$2.1\%, also significantly greater
than the transit depth due to the planetary disk alone ({\em Ben-Jaffel and Sona Hosseini}, 2010). Recently, two other observations
of extended upper atmospheres due to Lyman-$\alpha$ absorption during the transits of the short periodic gas giant, HD 189733b,
have also been reported ({\em Lecavelier des Etangs et al.}, 2010, 2012).
{\em Holmstr\"{o}m et al.} (2008), {\em Ekenb\"{a}ck et al.} (2010) and {\em Lammer et al.} (2011a) have demonstrated that the exosphere-magnetosphere
environment of hydrogen-rich gas giants in orbit locations $\leq$0.05 AU should be strongly affected by the production
of energetic neutral H atoms (ENAs) as well as non-thermal ion escape processes.

Similar processes can be expected for EUV-heated and hydrodynamically expanding upper atmospheres of terrestrial exoplanets,
where the upper atmosphere can expand beyond possible
magnetopause configurations so that ENAs will be produced via charge exchange
between the charged stellar wind plasma flow and the exospheric particles.
From the ionization of exospheric planetary neutral atoms one can probe of the stellar wind induced ion pick-up loss rate.
Recently, {\em Kislyakova et al.} (2013) applied a stellar wind plasma flow and upper atmosphere interaction model to hydrogen-rich terrestrial planets,
exposed to an EUV flux 100 times the present solar strength  within the habitable zone of a typical M-star at
 $\sim$0.24 AU. This study found  non-thermal escape rates of $\sim 5\times 10^{30}$~s$^{-1}$, which is about one order of magnitude
 weaker than the EUV-driven thermal escape rates.

\begin{figure}[t]
\includegraphics[width=\columnwidth]{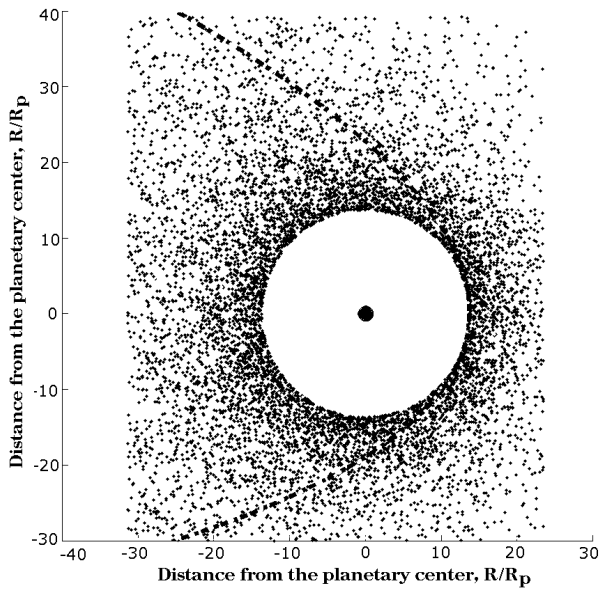}
\caption{\small 
Modeled extended neutral atomic hydrogen exosphere around a hydrogen-rich ``super-Earth''
inside an M star HZ at 0.24~AU with  the size of $2R_{\rm \bigoplus}$ and $10M_{\rm \bigoplus}$;
the EUV flux is is 50 times the present Sun's. The stellar wind plasma density and 
velocity are $\sim$250~cm$^{-3}$ and 330~km~s$^{-1}$, respectively.
The small dots represent the neutral hydrogen atoms, the big dot in the center 
represents the planet (after {\em Kislyakova et al.}, 2013).}
\label{fig:superearth}
\end{figure}

As one can see from Fig.~\ref{fig:superearth} the hydrogen atoms populate a wide area around the planet.
The EUV flux is deposited mainly in the lower thermosphere, while the fraction of the ENAs
which are directed towards the planet should be deposited in an atmospheric layer below the exobase
level where they contribute to thermospheric heating. The production of such huge hydrogen coronae
and related ENAs was predicted by {\em Chassefi\`{e}re} (1996a) when he studied the hydrodynamic escape
of hydrogen from a hot H$_2$O-rich early Venus thermosphere. This author suggested that the production
of ENAs under such extreme conditions should make an important contribution to the heat budget
in the upper atmosphere. The expected energy deposition to hydrodynamically expanding but adiabatically cooling
upper atmospheres has to be studied in the future because this important stellar wind-plasma induced feedback
process can rise the temperature to higher values than the EUV radiation, thus possibly enhancing
thermal escape.

% Sect. 7
\section{\textbf{APPLICATIONS}}

\bigskip
\noindent
\textbf{7.1 The Faint Young Sun Paradox}
\bigskip

Stellar evolution theory indicates that the ZAMS Sun was fainter by about 30\%
than the present-day Sun; the radiative output has increased gradually since
then (Sect. 4.1). Estimates of the Earth's and Mars' surface temperatures in the presence
of their modern atmospheres indicate that both should have been at average
temperatures much below the freezing point of liquid water ({\em Sagan and Mullen}, 1972).
Geological evidence, however, clearly suggests warm climates in the first
billion years, and the presence of liquid water ({\em Kasting and Toon}, 1989;
{\em Valley et al.}, 2002; {\em Feulner}, 2012). Without some source of additional
warming, the Earth's surface temperature should not have
risen above the freezing point until about 2 billion years ago, long after the
first traces of life on Earth appeared ({\em Kasting and Catling}, 2003).
This apparent contradiction is known as the ``Faint Young Sun Paradox'' (FYSP).

The FYSP has several different possible solutions, and it is not yet agreed which,
if  any of these, is correct. These solutions fall into three basic categories:

$\bullet$ {\em Additional greenhouse gases.} Higher levels of greenhouse gases
could raise the atmospheric temperatures significantly (important also in the
present-day Earth's atmosphere). For the Earth, increased levels of atmospheric
CO$_2$ together with water vapor would potentially suffice, and the CO$_2$ level
could even self-regulate the mild climate though the carbonate-silicate cycle
({\em Walker et al.}, 1981). However, geological evidence from paleosols ({\em Sheldon}, 2006;
{\em Driese et al.}, 2011) argues against massively increased CO$_2$ levels on the young
Earth.

Previous paleosol constraints published by {\em Rye et al.} (1995) have been
criticized ({\em Sheldon}, 2006), as have the very low CO$_2$
estimates from banded iron-formations published by {\em Rosing et al.} (2010)
({\em Reinhard and Planavsky}, 2011; {\em Dauphas and Kasting},
2011). The {\em Sheldon} and {\em Driese et al.} CO$_2$ estimates should be
regarded as highly uncertain, as they depend on poorly known factors such as
soil porosity and moisture content. Furthermore, recent 3-D climate calculations
 by {\em Kienert et al.} (2012) suggest that CO$_2$ partial pressures on the
 early Earth must have been even higher than previously calculated (see also in
 1D models {\em Haqq-Misra et al.}, 2008; {\em von Paris et al.}, 2008), because
 ice albedo feedback tends to amplify the cold temperatures predicted when
 CO$_2$ is low. All of this suggests that additional greenhouse gases were probably needed.

Other greenhouse gases such as NH$_3$ may dissociate rapidly ({\em Kuhn and
Atreya}, 1979), and CH$_4$ may form hazes, increasing the albedo, unless
at least an equal amount of CO$_2$ is present ({\em Haqq-Misra et al.}, 2008).
Ethane (C$_2$H$_6$) and carbonyl sulfide (OCS) have been proposed as
efficient greenhouse gases in the young terrestrial atmosphere ({\em Haqq-Misra
et al.}, 2008; {\em Ueno et al.}, 2009). Another candidate is N$_2$O which
can build up biotically to quite large concentrations and act as a strong greenhouse gas
({\em Buick}, 2007; {\em Roberson et al.}, 2011; {\em Grenfell et al.}, 2011).

Calculations by {\em Ramirez et al.} (2014) suggest
that the early Mars climate problem could instead be resolved by adding
5-20\% H$_2$ to the atmosphere, as has been suggested recently for early Earth
({\em Wordsworth and Pierrehumbert}, 2013).  In addition, higher partial
pressures of N$_2$ might enhance the effectiveness of the CO$_2$ greenhouse effect, on
both early Earth and early Mars ({\em Li et al.}, 2009; {\em Goldblatt et al.}, 2009;
{\em von Paris et al.}, 2013). Furthermore, SO$_2$ has been investigated for early Mars
climate studies ({\em Postawko and Kuhn}, 1986; {\em Tian et al.}, 2010; {\em Mischna et al.},
2013).

$\bullet$ {\em Lower albedo.} A planet's albedo could be changed by lower
cloud coverage in particular in cooler environments ({\em Rossow et al.}, 1982).
Detailed studies of clouds show two competing effects ({\em Goldblatt \& Zahnle}, 2011a;
{\em Zsom et al.}, 2012):
Low-lying clouds reflect solar radiation, thus increasing the albedo; in contrast,
high clouds, in particular ice-crystal cirrus clouds, add to atmospheric greenhouse
warming, potentially sufficient to solve the FYSP ({\em Rondanelli \& Lindzen}, 2010)
although requiring full coverage by an unrealistically thick and cold cloud cover
({\em Goldblatt \& Zahnle}, 2011a).

A lower surface albedo could be the result of the suggested smaller continental area in early
epochs although feedback on cloud coverage may result in opposite trends. The absence of
cloud condensation nuclei induced biologically ({\em Charlson et al.} 1987) also decreases the
albedo. Considering these effects and claiming less reflective clouds because of larger
cloud droplets (because of fewer biogenically induced condensation nuclei), {\em Rosing et al.}
(2010) inferred a clement climate for  the young Earth without need of high amounts of
greenhouse gases ({\em Rosing et al.}, 2010) although this solution was again questioned
({\em Goldblatt and Zahnle}, 2011b).

$\bullet$ {\em A brighter young Sun.} If the Sun had been slightly more
massive at arrival on the ZAMS, then its luminosity would have exceeded
standard estimates, owing to the $L \propto M^3$ dependence of a cool star's
luminosity, $L$, on the stellar mass, $M$. The excess mass could have been
shed by a much stronger solar wind in the early phases of solar evolution
({\em Whitmire et al.}, 1995; {\em Sackmann and Boothroyd}, 2003). This hypothesis
is constrained by the requirement that Martian temperatures be supportive of
liquid water, while preventing a runaway greenhouse on Earth.
The resulting ZAMS mass of the Sun would then be 1.03--1.07~$M_{\odot}$ ({\em Sackmann
and Boothroyd}, 2003). Upper limits to the ionized-wind mass-loss rates
deduced from radio observations of young solar analogs are marginally
compatible with the ``Bright Young Sun'' requirement, indicating a maximum
ZAMS solar mass of 1.06~$M_{\odot}$ ({\em Gaidos et al.}, 2000). Estimates of
the young Sun's wind mass-loss, however (Sect.~4.4), suggest a total
main-sequence mass-loss of the Sun of only 0.3\% ({\em Minton and Malhotra},
2007), too little to explain the FYSP.

\bigskip
\noindent
\textbf{7.2 Venus Versus Earth}
\bigskip

The atmosphere of present Venus is extremely dry when compared to the
terrestrial atmosphere (a total water content of $2 \times 10^{19}$~g
or 0.0014\% of the terrestrial ocean, vs.\ $1.39 \times
10^{24}$~g on Earth; {\em Kasting and Pollack}, 1983). The similar masses
of Venus and Earth and their formation at similar locations in the
solar nebula would suggest that both started with similar water
reservoirs ({\em Kasting and Pollack}, 1983). Terrestrial planets such as Venus, Earth and Mars,
with silicate mantles and metallic cores, most likely obtain water and
carbon compounds during their accretion (see Sect. 3). The largest part of a
planet's initial volatile inventory is outgassed into the atmosphere
during the end of magma ocean solidification (e.g., {\em Zahnle et al.}, 1988, 2007;
{\em Elkins-Tanton}, 2008; {\em Lebrun et al.}, 2013; {\em Hamano et al.}, 2013;
{\em Lammer}, 2013). Only a small fraction of the initial volatile inventory will be
released into the atmospheres through volcanism. From estimates
by considering the possible range of the initial water content in the building
blocks of Venus and Earth it appears possible that both planets
may have outgassed hot steam atmospheres with surface pressures
up to $\sim$500 bar ({\em Elkins-Tanton}, 2008).

Recently, {\em Lebrun et al.} (2013), studied the thermal evolution of
early magma oceans in interaction with outgassed steam atmospheres for
early Venus, Earth and Mars. Their results indicate that the water vapor of
these steam atmospheres condensed to an liquid ocean after $\sim$10, 1.5 and
0.1 Myr for Venus, Earth and Mars, respectively, and remained in vapor form for Earth-type
planets orbiting closer than 0.66 AU around a solar-like star. However, these
authors point out that their results depend on the chosen opacities and
it is conceivable  that no liquid
oceans formed on early Venus. This possibility is also suggested by
{\em Hamano et al.} (2013) who studied the emergence of steam atmosphere-produced
water oceans on terrestrial planets.
%---------------------Fig. Venus
\begin{figure}[t!]
\begin{center}
\includegraphics[width=0.5\textwidth, clip=]{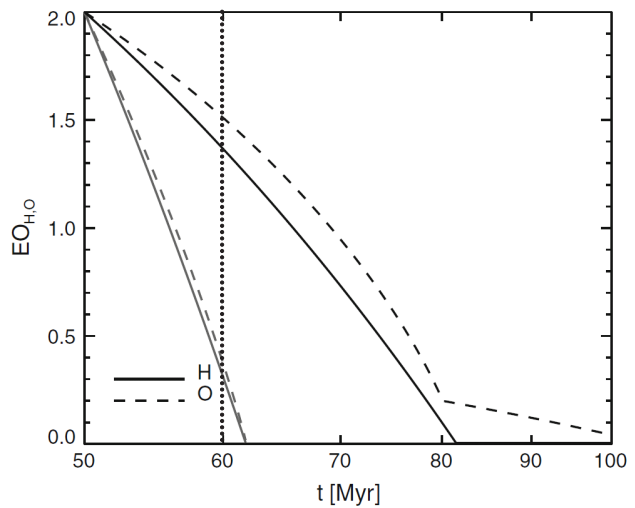}
\caption{\small Hydrodynamic loss of atomic hydrogen and dragged oxygen atoms from a 500 bar outgassed, dissociated steam
atmosphere on early Venus during the young Sun's EUV saturation phase (100 times the present 
EUV flux). The black and grey lines correspond to heating efficiencies of 15\% and 40\%, respectively (after {\em Lammer}, 2013). 
The dotted line marks the time  when the steam atmosphere may have condensed ({\em Lebrun et al.}, 2013) so 
that an ocean could have formed on the planet's surface.}
\label{fig:Venus}
\end{center}
\end{figure}
%----------------------

The critical solar flux
leading to complete evaporation of a water ocean (a ``runaway
greenhouse'', {\em Ingersoll}, 1969) is $\approx 1.4\,S_\mathrm{sun}$
nearly independently of the amount of cooling $\mathrm{CO}_2$ in the
atmosphere ({\em Kasting}, 1988), where $S_\mathrm{sun} = 1360$~Wm$^{-2}$
is the present-day solar constant. This critical flux
is just about the flux expected from the young Sun at the orbit of
Venus (and is about 70\% of the present-day value at the same solar
distance). New calculations by {\em Kopparapu et al.} (2013) give
a runaway greenhouse limit of $1.06\,S_\mathrm{sun}$. That limit, however,
like the {\em Kasting} (1988) value before it, is almost certainly too low,
because it assumes a fully saturated atmosphere and it neglects cloud feedback.

Venus' atmosphere also reveals a
surprisingly high deuterium-to-hydrogen (D/H) abundance ratio of $(1.6
\pm 0.2) \times 10^{-2}$ or 100~times the terrestrial value of $1.56
\times 10^{-4}$ ({\em Kasting and Pollack}, 1983). These observations have
motivated a basic model in which water was abundantly available on the
primitive Venus. Any water oceans would evaporate, inducing a
``runaway greenhouse'' in which water vapor would become the major
constituent of the atmosphere. This vapor was
then photodissociated in the upper atmosphere by {\it enhanced} solar
EUV irradiation ({\em Kasting and Pollack}, 1983) as well as by frequent meteorite impacts in
the lower thermosphere ({\em Chassefi\`{e}re}, 1996b), followed by the escape of hydrogen
into space ({\em Watson et al.}, 1981; {\em Kasting and Pollack}, 1983; {\em Chassefi\`{e}re}, 1996a, 1996b; {\em Lammer et al.}, 
2011c). During the phase of strong thermal escape of atomic hydrogen, remaining oxygen
and similar species can be dragged to space by the outward flowing hydrogen flux ({\em Hunten 
et al.}, 1987; {\em Zahnle and Kasting}, 1986; {\em Chassefi\`{e}re}, 1996a, 1996b; {\em Lammer et al.}, 2011c).

Fig.~\ref{fig:Venus}  shows the hydrodynamic escape of a hypothetical 500 bar steam atmosphere from 
early Venus exposed to an EUV flux 100 times the present solar value, for  assumed heating efficiencies 
of 15\% and 40\% ({\em Lammer et al.}, 2011c; {\em Lammer}, 
2013). In this scenario, it is assumed that early Venus finished its accretion $\sim$50 Myr after the
origin of the Sun and outgassed the maximum expected fraction  of its initial water inventory
after the magma ocean solidified. One can see that the whole steam atmosphere could have been lost 
within the condensation time scale of about 10 Myr modeled by {\em Lebrun et al.} (2013) for
a heating efficiency of 40\% as estimated for ``Hot Jupiters'' ({\em Koskinen et al.}, 2013). 
For a heating efficiency of 15\% (e.g., {\em Chassefi\`{e}re}, 1996a, 1996b) and assuming
that the outgassed steam atmosphere remains in vapor form, the planet would have 
lost its hydrogen  after $\sim$30 Myr, leaving behind some of its oxygen. If
the remaining O atoms also experienced high thermal escape rates during the remaining high EUV phase, 
they would be lost by a combination of thermal and non-thermal escape processes ({\em Lundin et al.}, 2007).
If the high-density CO$_2$ lower thermosphere was efficiently cooled down to suppress high thermal escape rates of oxygen, 
the remaining oxygen in the upper atmosphere could have been lost by non-thermal escape processes such as solar wind 
induced ion pick-up and cool ion outflow (e.g., {\em Kulikov  et al.}, 2006; {\em Lundin  et al.}, 
2007; {\em Lammer}, 2013).

If the steam atmosphere cooled within $\sim$10 Myr, so that the remaining water vapor could 
condense, then a liquid water ocean may have formed on  Venus' surface for some period before it 
again evaporated by the above mentioned ``runaway greenhouse'' effect.
On the other hand, if early Venus degassed an initial water content that resulted in a steam atmosphere 
with $<250$ bar, then the water would most likely have been lost during its vapor phase so that an ocean 
could never form on the planet's surface.

As discussed in {\em Lammer et al.} (2011c) and studied in detail in {\em Lebrun et al.} (2013) 
and {\em Hamano et al.} (2013), the outgassed steam atmosphere of the early Earth cooled faster because 
of its larger distance from the Sun, and this resulted in a shorter condensation time scale compared to the escape. 
Therefore, a huge fraction of the water vapor condensed and produced Earth's early ocean.

\bigskip
\noindent
\textbf{7.3 Mars}
\bigskip

\begin{figure}[t!]
\begin{center}
\includegraphics[width=0.5\textwidth, clip=]{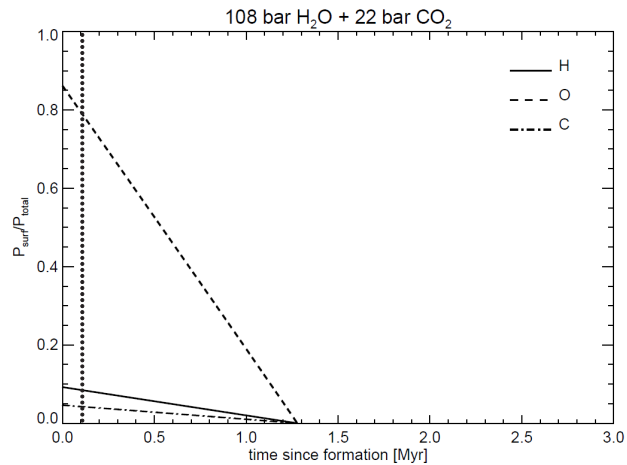}
\caption{\small Hydrodynamic loss of atomic hydrogen and dragged oxygen and carbon atoms from an outgassed 
and dissociated steam atmosphere (108 bar H$_2$O, 22 bar CO$_2$) on early Mars during the young Sun's EUV saturation 
phase (100 times the present solar EUV level; after 
{\em Erkaev et al.}, 2014). The dotted line marks the time period when the steam atmosphere may have condensed 
({\em Lebrun et al.}, 2013), so that an ocean could have formed on the planet's surface (neglecting the energy  
deposited at the planet's surface by frequent planetesimal impacts and small  embryos).}
\label{fig:Mars}
\end{center}
\end{figure}

Latest research on the formation of Mars indicates that the body formed within $\sim$5 Myr and 
remained as a planetary embryo that never accreted toward a
larger planet ({\em Brasser}, 2013). From the building blocks and their water contents
expected in the Martian orbit, it is estimated that early Mars catastrophically outgassed volatiles amounting
to $\sim$50--250 bar of H$_2$O and $\sim$10--55 bar of CO$_2$,
after the solidification of its magma ocean ({\em Elkins-Tanton}, 2008). As shown in Fig.~\ref{fig:Mars},
due to the high EUV flux of the young Sun, such a steam atmosphere was most likely lost via hydrodynamic 
escape of atomic hydrogen, dragging heavier atoms such as C and O during the first million years after 
the magma ocean solidified ({\em Lammer et al.}, 2013c; {\em Erkaev  et al.}, 2014). Although {\em Lebrun  
et al.} (2013) found that a degassed steam atmosphere on early Mars would condense after $\sim$0.1 Myr, 
frequent impacts of large planetesimals and small embryos during the early Noachian most likely
kept the protoatmosphere in vapor form ({\em Lammer et al.}, 2013c). After early Mars lost its outgassed 
protoatmosphere the atmospheric escape rates were most likely balanced by
a secondary outgassed atmosphere and delivered volatiles by impacts
until the activity of the young Sun decreased so that the atmospheric sources could dominate over the 
losses ({\em Tian et al.}, 2009; {\em Lammer et al.}, 2013c).
%---------------------Fig. Mars
%----------------------
There is substantial evidence for a warmer and wetter climate on
ancient Mars $\sim$3.5--4 Gyr ago ({e.g., {\em Carr and Head}, 2003)}.
Mantle outgassing and impact-related volatiles may have resulted in
the buildup of a secondary CO$_2$ atmosphere of a few tens to a few
hundred mbar around the end of the Noachian ({\em Grott et al.}, 2011; {\em Lammer  et al.}, 2013c). The 
evolution of the planet's secondary atmosphere and its water content during the past 4 Gyr was most likely
determined by an interplay of non-thermal atmospheric escape processes such as direct escape of photochemically 
hot atoms, atmospheric sputtering, ion pick-up and cool ion outflow as well as impacts (e.g., {\em Lundin et al.}, 
2007; {\em Lammer et al.} 2013c and references therein), carbonate precipitation, and serpentinization 
({\em Chassefi\`{e}re and Leblanc}, 2011a, 2011b; {\em Chassefi\`{e}re et al.}, 2013) which  finally led 
to the present-day environmental conditions.

% Sect. 8
\section{\textbf{THE VARIETY OF HABITABLE CONDITIONS}}

With the rapidly increasing number of detected exoplanetary systems, the diversity of system
architectures, sizes and radii of the individual planets and the range of average planetary
densities have come into the focus of {\it planetary characterization}.  It is by now clear
that a much wider range of planetary characteristics is realized in exoplanetary systems than
available in the solar system. Hot Jupiters, massive planets in habitable zones, resonant systems,
planets in highly eccentric orbits, and small planets with very low average densities (below that of water) 
are challenging planetary formation, migration, and evolution theories.

Whether a planet is ultimately habitable depends on a surprisingly large number of astrophysical
and geophysical factors, of which we have reviewed the former here. The correct distance from the host
star is the most obvious and defining parameter for a habitable zone, but it provides by no means
a sufficient condition for habitability. Migration and resonant processes in the young planetary systems are
crucial for the fate of stable orbits of potentially habitable planets. In what looks like a tricky
paradox, water is not available in solids in the very region where habitable zones evolve around cool stars; rather,
water needs to be brought there, and gas giant planets may have to do the job of stirring the orbits of the
outer planetesimals to scatter them to regions where water is scarce. Water transport through collisions is
itself a difficult process, as water needs to survive such collisions to accumulate in the required
amounts on the forming planet. Atmospheres outgassed from the planet and trapped from the early 
protoplanetary disk wrap the growing planet into an extremely dense envelope that itself defies
habitability but may be protective of forming secondary atmospheres. Once atmospheres evolve,
the harsh environment of the young star begins to act severely. EUV and X-ray radiation hundreds to thousands of 
times stronger than the present-day Sun's and stellar winds up to hundreds of times stronger than at present
begin to erode upper atmospheres of planets, in particular if no magnetosphere is present that may
sufficiently protect the atmosphere. Ultraviolet emission determines the atmospheric temperature profiles
and induces a rich chemistry that may become relevant for the formation of life. Here, host star spectral type
matters for the formation of an atmospheric temperature inversion. An abiotically forming
ozone layer may be one of the important results especially for active M dwarfs.

The outcome may be an extremely rich variety of potentially habitable environments, but we still lack
thorough knowledge of all relevant factors that really determine habitability. While M dwarfs have
become favored targets to search for planets in their habitable zones, such planets may be affected
by the strong, long-term high-energy radiation and harsh wind conditions for too long to actually develop
habitable planetary surfaces. Their atmospheres may be subject to excessively strong erosion processes.

On the other hand, the development of two habitable planets in the solar system, at least one with thriving
life, provides ample evidence that some of the risks, e.g., too heavy protoatmospheres or too rapid erosion 
of early nitrogen atmospheres, were successfully overcome ({\em Lammer et al.}, 2013a). One wonders
if some sort of fine-tuning was at work, for  example between the timing of planet formation 
and disk dispersal, the outgassing history and the eroding high-energy radiation, the presence of giants 
like Jupiter or Saturn controlling the architecture and stability of planetary orbits in the habitable zone,
or the mass- and perhaps rotation-dependent formation of a magnetosphere protecting the upper atmosphere
of a planet. If such did not succeed, then habitable-zone planets may end up with huge high-pressure hydrogen
atmospheres, suggested by the detection of small, low-density exoplanets, or strongly eroded atmospheres 
such as on Mars, or the complete loss of their atmospheres. Other planets may be close to habitable zones
but may dwell in eccentric orbits, making surface life more problematic.

Apart from the diverse astrophysical conditions, a plethora of {\em geophysical} conditions matter,
such as planetary internal convection and geodynamos, plate tectonics, volcanism, the presence and distribution 
of land mass, the orientation and stability of the rotation axis, etc. They as well interact with the
diverse astrophysical factors discussed in this article, to add further constraints on actual habitability.

\medskip

\textbf{ Acknowledgments.} We thank an anonymous referee for useful comments that helped improve this chapter.
M. G., R. D., M. L. K., H. L. and E. P.-L. thank the Austrian Science Fund (FWF) for 
financial support (research grants related to planetary habitability: S116001-N16 and S116004-N16 for M. G.,
S11603-N16 for R. D., S116606-N16 for M. L. K., S116607-N16 for H. L., S11608-N16 and P22603-N16 for E. P.-L.). 
N. V. E.  acknowledges support by the RFBR grant No 12-05-00152-a. J. K. thanks the NASA Astrobiology 
and Exobiology programs. H. R. and H. L. thank for the support by the Helmholtz Association through the 
research alliance ``Planetary Evolution and Life." 
I. R. acknowledges financial support from the Spanish Ministry of Economy and Competitiveness 
(MINECO) and the "Fondo Europeo de Desarrollo Regional" (FEDER) through grant AYA2012-39612-C03-01. M. G., H. L., 
and M. L. K. acknowledge support by the International Space Science Institute in Bern, Switzerland, and the
ISSI Team "Characterizing Stellar and Exoplanetary Environments."

\bigskip
\newpage

\centerline\textbf{ REFERENCES}
\bigskip
\parskip=0pt
{\small
\baselineskip=11pt

 \refs Abe Y., et al. (2011) {\em Astrobiol., 11}, 443-460.

 \refs Adams E. R., et al. (2008) {\em Astrophys. J., 673}, 1160-1164.   % changed July 3, MG

 \refs Albar\`{e}de F. and Blichert-Toft J. (2007) {\em C. R. Geosci., 339}, 917-927.

 \refs Anglada-Escud\'e G.,  et al. (2012) {\em Astrophys. J., 751}, id.L16.
 
 \refs Audard  M., et al. (2000) {\em Astrophys. J., 541}, 396-409.

 \refs Baraffe I., et al. (1998) {\em Astron. Astrophys., 337}, 403-412.
 
 \refs Baross J. A., et al. (2007) In {\it Planets and Life: The Emerging Science of Astrobiology}
      (W. T. I. Sullivan and J. A. Baross, eds) pp. 275-291. Cambridge Univ. Press, Cambridge.

 \refs Batalha N. M., et al. (2011) {\em Astrophys. J., 729}, id.27.

 \refs Batalha N. M., et al. (2013) {\em Astrophys. J. Suppl. Ser., 204}, id.24.

 \refs Baumjohann W. and Treumann R. A. (1997) {\em Basic space plasma physics}, Imperial College Press, London.

 \refs Ben-Jaffel L. (2007) {\em ApJ, 671}, L61-L64.

 \refs Ben-Jaffel L. and Sona Hosseini S. (2010) {\em ApJ, 709}, 1284-1296.

 \refs Bonfils X., Delfosse X., Udry S., et al. (2013) {\em Astron. Astrophys., 549}, A109.
 
 \refs Borucki W. J., et al. (2013) {\em Science, 340}, 587-590.
 
 \refs Brasser R. (2013) {\em Space Sci. Rev., 174}, 11-25.

 \refs Buchhave L. A., et al. (2012) {\em Nature, 486}, 375-377.

 \refs Buick R. (2007) {\em Geobiol., 5}, 97-100.

 \refs Carr M. H. and Head J. W. (2003)  {\em  J. Geophys. Res., 108}, 8/1-28, CiteID 5042.

 \refs Chamberlain J. W. (1963) {\em Planet. Space. Sci., 11}, 901-996.

 \refs Chapman S. A. (1930) {\em Mem. R. Met. Soc., 3}, 103-125.

 \refs Charlson R. J., et al. (1987) {\em Nature, 326}, 655-661.

 \refs Chassefi\`{e}re E. (1996a) {\em J. Geophys. Res., 101}, 26039-26056.

 \refs Chassefi{\`e}re E. (1996b) {\em Icarus, 124}, 537-552.

 \refs Chassefi\`{e}re E. and Leblanc F. (2011a) {\em Earth Planet. Sci. Lett., 310}, 262-271.

 \refs Chassefi\`{e}re E. and Leblanc F. (2011b) {\em Planet. Space Sci. 59}, 218-226.

 \refs Chassefi\`{e}re E., et al. (2013) {\em J. Geophys. Res. 118}, 1123-1134.

 \refs Chiang E. I. (2003) {\em Astrophys. J., 584}, 465-471.

 \refs Christensen U. R. (2010) {\em Space Sci. Rev., 152}, 565-590.

 \refs Christensen U. R. and Aubert J. (2006) {\em Geophys. J. Int., 166}, 97-114.

 \refs Claire M. W., et al. (2012) {\em Astrophys. J., 757}, id. 95.

 \refs Crutzen P. J. (1970) {\em Q. J. R. Meteorol. Soc., 96}, 320-325.

 \refs Dauphas N. and Kasting J. F. (2011) {\em Nature, 474}, E2-E3.

 \refs Drake S. A., et al. (1993) {\em Astrophys. J.,  406}, 247-251.

 \refs Driese S. G., et al. (2011) {\em Precambrian Res., 189}, 1-17.

 \refs Dvorak R.,  et al. (2012), in {\em Let's Face Chaos through Nonlinear Dynamics} 
      (M. Robnik and V. G. Romanovski, eds.), pp. 137-147. American Institute of Physics, Melville, NY.

 \refs Eggl S., et al. (2012) {\em Astrophys. J., 752}, 74-85. 

 \refs Ekenb\"{a}ck A., et al., (2010) {\em Astrophys. J., 709}, 670-679.

 \refs Elkins-Tanton L. T. (2008) {\em Earth Planet. Sci. Lett. 271}, 181-191.

 \refs Elkins-Tanton L. T. (2011) {\em Astrophys. Space Sci., 332}, 359-364.

 \refs Elkins-Tanton L. and Seager S. (2008) {\em Astrophys. J., 685}, 1237-1246.

 \refs Erkaev N. V.,  et al. (2005) {\em Astrophys. J. Suppl. Ser., 157}, 396-401.

 \refs Erkaev N. V.,  et al. (2013) {\em Astrobiol., 13}, 1011-1029.
 
 \refs Erkaev N. V.,  et al. (2014) {\em Planet. Space Sci., 98},  106-119.    ; changed MG 29 July 2014

 \refs Farrell W. M., et al. (1999) {\em  J. Geophys. Res., 104}, (E6), 14025-14032.

 \refs Feldman W. C., et al. (1977) In {\em The Solar Output and its
      Variation}, (O. R. White, ed.), pp. 351-382. Colorado Associated University Press.

 \refs Feulner G. (2012) {\em Reviews of Geophysics, 50}, CiteID RG2006.

 \refs Forget F. and Pierrehumbert R. T. (1997) {\em Science, 278}, 1273-1276.

  \refs France K., et al. (2013) {\em Astrophys. J., 763}, id. 149.
 
 \refs Fressin F., et al. (2013) {\em Astrophys. J., 766}, id.81.
 
 \refs Garc{\'e}s A., et al. (2011) {\em Astron. Astrophys., 531}, A7.

 \refs Gaidos E. J., et al. (2000) {\em Geophys. Res. Lett.,  27}, 501-503.

 \refs Gayley K. G., et al. (1997) {\em Astrophys. J. 487}, 259-270.

 \refs Gilliland R. L., et al.  (2011) {\em Astrophys. J. Suppl. Series, 197}, id.6.

 \refs Goldblatt M. T. and Zahnle  K.J. (2011a)  {\em Clim. Past, 7}, 203-220.

 \refs Goldblatt M. T. and Zahnle K. J. (2011b) {\em Nature, 474}, E3-4.

 \refs Goldblatt C., et al. (2009), {\em Nature Geoscience, 2}, 891-896.

 \refs Grenfell J. L.,   et al. (2007) {\em Astrobiol., 7}, 208-221.
 
 \refs Grenfell J. L.,  et al. (2011) {\em Icarus, 211}, 81-88.

 \refs Grenfell J. L.,  et al. (2012) {\em Astrobiol., 12}, 1109-1122.

 \refs Grenfell J. L., et al. (2013) {\em Astrobiol., 13}, 415-438. 

 \refs Grenfell J. L.,  et al. (2014) {\em Planet. Space Sci., 98}, 66-76.  ; added MG 29 July 2014
      
 \refs Grie{\ss}meier J.-M.,  et al. (2004) {\em Astron. Astrophys., 425}, 753-762.

 \refs Grott M., et al. (2011) {\em Earth Planet. Sci. Lett. 308}, 391-400.

 \refs G\"udel M. (2004) {\em Astron. Astrophys. Rev., 12}, 71-237.

 \refs G\"udel M. (2007) {\em Liv. Rev. Solar Phys., 4}, no. 3; online
      http://solarphysics.livingreviews.org/Articles/lrsp-2007-3/.

 \refs G{\"u}del M., et al. (1997) {\em Astrophys. J. 483}, 947-960.

 \refs Guinan E. F. and Engle S. G. (2009) In  {\it The Ages of Stars} (E. E. Mamajek, D. R. Soderblom, and R. F. G. Wyse, eds.)
      {\em IAU Symp. 258}, 395-408.
 
 \refs Haagen-Smit A. J., et al. (1952) {\em 2nd Proc. Nat. Air. Symp.}, 54-56. 

 \refs Haghighipour N. and  Kaltenegger L. (2013) {\em Astrophys. J., 777}, id.166.

 \refs Hamano K., et al. (2013) {\em Nature, 497}, 607-610.

 \refs Haqq-Misra J. D., et al.  (2008) {\em Astrobiol., 8}, 1127-1137.

 \refs Hart M. H. (1979) {\em Icarus, 37}, 351-357.

 \refs Havnes O. and Goertz C. K. (1984) {\em Astron. Astrophys., 138}, 421-430.

 \refs Hayashi C., et al. (1979) {\em Earth \&  Planet. Sci. Lett.,  43}, 22-28. 

 \refs Hedelt P.,  et al. (2013) {\em Astron. Astrophys., 553}, idA9. 

 \refs Holland H. D. (1962) In {\em Petrologic studies} (A. E. J. Engel, H. L. James, and B. F. Leonard, eds.)
      Geol. Soc. Amer., Boulder, Colorado.

 \refs Holmstr\"{o}m M.,  et al. (2008) {\em Nature, 451}, 970-972.
 
 \refs Howard A. W. (2013)  {\em Science, 340}, 572-576.
 
 \refs Hu R., et al. (2012) {\em Astrophys. J., 761}, id. 166.

 \refs Hunten D. M., et al. (1987) {\em Icarus, 69}, 532-549.

 \refs Ikoma M., et al. (2000) {\em Astrophys. J., 537}, 1013-1025.

 \refs Ikoma M. and Genda H. (2006) {\em Astrophys. J.}, 648, 696-706.

 \refs Ingersoll A. P. (1969) {\em J. Atmospheric Sci., 26}, 1191-1198.

 \refs Ip W.-H., et al. (2004) {\em Astrophys. J., 602}, L53-L56.

 \refs Johansson E. P. G., et al. (2009) {\em Astron. Astrophys., 496}, 869-877.

 \refs Kaltenegger L. and Haghighipour N. (2013) {\em Astrophys. J., 777}, id.165.

 \refs Kaltenegger L., et al. (2007) {\em Astrophys. J., 658}, 598-616. 

 \refs Kaltenegger L., et al. (2011) {\em Astrophys. J., 733}, id.35. 

 \refs Kaltenegger L., et al. (2013) {\em Astrophys. J., 755}, id.L47.

 \refs Kane S. R. and Hinkel N. R. (2013) {\em Astrophys. J., 762}, id.7.
 
 \refs Kasting J. F. (1988) {\em Icarus, 74}, 472-494.

 \refs Kasting J. F. and Catling D. (2003) {\em Annu. Rev. Astron. Astrophys., 41}, 429-463.

 \refs Kasting J. F. and Pollack J. B. (1983) {\em Icarus, 53}, 479-508.

 \refs Kasting J. F. and Toon O. B. (1989)
      In {\em Origin and Evolution of Planetary and Satellite Atmospheres}, (S. K. Atreya, J. B.
      Pollack, and M. S. Matthews, eds), pp. 423-449, University of Arizona Press, Tucson, USA.

 \refs Kasting J. F., et al. (1993) {\em Icarus, 101}, 108-128.
 
 \refs Kasting J. F., et al. (2014) {\em Proc. Nat. Acad. Sci.}, in press.

 \refs Kharecha P., et al. (2005) {\em Geobiol., 3}, 53-76.

 \refs Khodachenko M. L., et al. (2007a) {\em Astrobiol., 7}, 167-184.

 \refs Khodachenko M. L., et al. (2007b) {\em Planet. Space Sci., 55}, 631-642.

 \refs Khodachenko M. L.,  et al.  (2012) {\em Astrophys. J., 744}, 70-86.

 \refs Kienert H., et al. (2012) {\em Geophys. Res. Lett. 39}, Issue 23, CiteID L23710.

 \refs Kipping D. M., et al. (2013) {\em Mon. Not. Roy. Astron. Soc., 434}, 1883-1888.

 \refs Kislyakova G. K.,  et al. (2013) {\em Astrobiol., 13}, 1030-1048.

 \refs Kitzmann D., et al. (2010) {\em Astron. Astrophys., 511}, id.A66.

 \refs Kitzmann D., et al. (2011a) {\em Astron. Astrophys., 531}, id.A62.

 \refs Kitzmann D., et al. (2011b) {\em Astron. Astrophys., 534}, id.A63.

 \refs Kitzmann D., et al. (2013) {\em Astron. Astrophys., 557}, id.A6.

 \refs Kopparapu R. K. (2013) {\em Astrophys. J., 767}, id.L8.

 \refs Kopparapu R. K., et al. (2013) {\em Astrophys. J., 765}, id.131.

 \refs Koskinen T. T., et al. (2007) {\em Nature, 450}, 845-848.

 \refs Koskinen T., et al. (2010) {\em Astrophys. J., 723}, 116-128.

 \refs Koskinen T. T., et al. (2013) {\em Icarus, 226}, 1678-1694.   % all ok now?

 \refs Kuhn W. R. and Atreya S. K. (1979) {\em Icarus, 37}, 207-213.

 \refs Kulikov Y. N., et al. (2006) {\em Planet. Space Sci., 54}, 1425-1444.

 \refs Kulikov Yu. N., et al. (2007) {\em Space Sci Rev., 129}, 207-243.

 \refs Lammer H. (2013) {\em Origin and evolution of planetary atmospheres: Implications for habitability},
      Springer Briefs in Astronomy, Springer, pp. 98, Heidelberg/New York.
      
 \refs Lammer H.,  et al. (2009) {\em Astron. Astrophys., 506}, 399-410.

 \refs Lammer H.,  et al., (2011a) {\em Astrophys. Space Sci., 335}, 9-23.

 \refs Lammer H., et al. (2011b) {\em Astrophys. Space Sci., 335}, 39-50. 

 \refs Lammer H., et al. (2011c) {\em Orig. Life Evol. Bisoph. 41}, 503-522.

 \refs Lammer H.,  et al. (2013a) In: {\em The early evolution of the atmospheres of terrestrial
      planets} (J. M. Trigo-Rodiguez, F. Raulin, F., C. Muller, and C. Nixon) {\em Astrophys. and Space Science Proc.}, pp. 35-52. Springer Verlag,
      Heidelberg, New York.

 \refs Lammer H.,  et al. (2013b) {\em Mon. Not. Roy. Astron. Soc., 430}, 1247-1256.

 \refs Lammer H.,  et al. (2013c) {\em Space Sci. Rev. 174}, 113-154.

 \refs Lebrun T.,  et al. (2013) {\em J. Geophys. Res., 118}, 1-22. % doi:10.1002/jgre.20068

 \refs Lecavelier des Etangs A.,  et al., (2010) {\em Astron. Astrophys., 514}, A72.

 \refs Lecavelier des Etangs A., et al. (2012) {\em Astron. Astrophys., 543}, L4.

 \refs Leconte J., et al. (2013) {\em Astron. Astrophys., 554}, A69.

 \refs Lederberg J. (1965) {\em Nature, 207}, 9-13.

 \refs L\'eger A., et al. (2009) {\em Astron. Astrophys., 506}, 287-302.
 
 \refs Li K., Pahlevan K., et al. (2009) {\em Proc. Nat. Academy Sci.}, 9576-9579.

 \refs Lichtenegger H. I. M.,  et al. (2009) {\em Geophys. Res. Lett. 36}, CiteID L10204.

 \refs Lim J. and White S. M. (1996) {\em Astrophys. J., 462}, L91-L94.

 \refs Lim J., et al. (1996)  {\em Astrophys. J., 460}, 976-983.

 \refs Linsky J. L. and Wood B. E. (1996) {\em Astrophys. J., 463}, 254-270.

 \refs Linsky J. L., et al. (2012) {\em Astrophys. J., 745}, id. 25.

 \refs Linsky J. L., et al. (2013) {\em Astrophys. J., 766}, id. 69.

 \refs Lissauer J. J. and the Kepler Team (2011) {\em Nature, 470}, 53-58.
 
 \refs Lovelock J. E. (1965) {\em Nature, 207}, 568-570.

 \refs Lundin R., et al. (2007) {\em Space Sci. Rev., 129}, 245-278.
 
 \refs Lunine J.,  et al.  (2011) {\em Adv. Sci. Lett., 4}, 325-338. 

 \refs Maehara H.,  et al. (2012) {\em Nature, 485}, 478-481.

 \refs Maindl T. I. and Dvorak, R. (2014), in {\em IAU Symposium 299: Exploring the
       formation and evolution of planetary systems} (B. Matthews and J. Graham, eds.), pp. 370-373,
       International Astronomical Union.     ; added MG 29 July 2014
       
 \refs Maindl T. I.,  et al. (2013) {\em Astron. Notes, 334}, 996.

 \refs Mamajek E. E. and Hillenbrand L. A. (2008) {\em Astrophys. J.,  687}, 1264-1293.

 \refs Marley M. S., et al. (2013)
      In: {\em Comparative Climatology of Terrestrial Planets} (S. Mackwell,
      M. Bullock, and J. Harder, eds.), University of Arizona Press, Tucson, AZ.

 \refs Mestel L. (1968), {\em Mon. Not. Roy. Astron. Soc., 138}, 359-391.

 \refs Minton D. A. and Malhotra R. (2007) {\em Astrophys. J., 660}, 1700-1706.

 \refs Mischna M. M., et al. (2000) {\em Icarus, 145}, 546-554.
 
 \refs Mischna M. A.,  et al.  (2013) {\em J. Geophys. Res. (Planets), 118}, 560-576.

 \refs Mizuno H. (1980) {\em Prog. Theor. Phys., 64}, 544-557.

 \refs Mizuno H., et al. (1978) {\em Prog. Theor. Phys., 60}, 699-710.

 \refs Mont\'esi L. G. J. and Zuber M. T. (2003) {\em J. Geophys. Res.-Planets, 108}, 2/1-25, CiteID 5048.

 \refs Morbidelli A. (2013) In {\em Planets, Stars and Stellar Systems} (T. D. Oswalt, L. M. French, and P. Kalas), pp 63-109.
       Springer Science+Business Media, Dordrecht. 

 \refs Morbidelli A., et al. (2000) {\em Meteoritics \& Planet. Sci., 35}, 1309-1320.

 \refs Morbidelli A., et al. (2007) {\em Astron. J., 134}, 1790-1798.

 \refs Mordasini C., et al. (2012) {\em Astron. \& Astrophys., 547}, A112.

 \refs National Academy of Sciences (2007) {\em The Limits of Organic Life in Planetary Systems}, National Acad. Sci., Washington, DC.

 \refs O'Brien D. P., et al. (2006) {\em Icarus, 184}, 39-58.

 \refs Osten R. A., et al. (2010) {\em  Astrophys. J., 721}, 785-801.

 \refs Parker E. N. (1958) {\em Astrophys. J., 128}, 664-675.

 \refs Paynter D. J. and  Ramaswamy, V. (2011) {\em J. Geophys. Res., 116}, D20302.

 \refs Penz T.,  et al. (2008) {\em Planet. Space Sci., 56}, 1260-1272.

 \refs Pepe F., et al. (2011) {\em Astron. Astrophys.}, 534, id.A58.
 
 \refs Pierrehumbert R. and Gaidos E. (2011) {\em Astrophys. J., 734}, id L13.

 \refs Pilat-Lohinger E., et al. (2008) {\em Astrophys. J., 681}, 1639-1645.  

 \refs Pizzolato N., et al. (2003) {\em Astron. Astrophys., 397}, 147-157.

 \refs Pollack J. B., et al. (1996) {\em Icarus, 124}, 62-85.

 \refs Postawko S. E. and Kuhn W. R. (1986) {\em J. Geophys. Res., 91}, 431-438.

 \refs Preusse S., et al. (2005) {\em Astron. Astrophys., 434}, 1191-1200.

 \refs Rafikov R. R. (2006) {\em Astrophys. J., 648}, 666-682.

 \refs Ramirez R. M., et al. (2014) {\em Nature Geosci., 7}, 59-63.

 \refs Rauer H., et al. (2011)  {\em Astron. Astrophys., 529}, id.A8.

 \refs Raymond S. N., et al. (2004) {\em Icarus, 168}, 1-17.

 \refs Raymond S. N., et al. (2007) {\em Astrobiol., 7}, 66-84.
 
 \refs Raymond S. N., et al. (2009) {\em Icarus, 203}, 644-662.

 \refs Reiners A. and Christensen U. R. (2010) {\em Astron. Astrophys., 522}, idA13.

 \refs Reinhard C. T. and Planavsky N. J. (2011) {\em Nature, 474}, E1-E2.

 \refs Ribas I., et al. (2005) {\em Astrophys. J., 622}, 680-694.

 \refs Ribas I., et al. (2010) {\em Astrophys. J., 714}, 384-395.

 \refs Ringwood A. E. (1979) {\em Origin of the Earth and Moon}. Springer Publishing House,
      Heidelberg / New York.

 \refs Roberson A., et al. (2011) {\em Geobiol., 9}, 313-320.

 \refs Rondanelli R. and Lindzen R. S. (2010)  {\em J. Geophys. Res., 115}, D02108.

 \refs Rosing M. T., et al. (2010) {\em Nature, 464}, 744-747.

 \refs Ringwood A. E. (1979) {\em Origin of the Earth and Moon}. Springer Publishing House, Heidelberg / New York.

 \refs Rossow W. B., et al. (1982) {\em Science, 217}, 1245-1247.

 \refs Rugheimer S.,  et al. (2013) {\em Astrobiol., 13}, 251-269.

 \refs Rye R., et al. (1995) {\em Nature, 378}, 603-605.

 \refs Sackmann I.-J. and Boothroyd A. I. (2003) {\em Astrophys. J., 583}, 1024-1039.

 \refs Sagan C. and Mullen G. (1972) {\em Science, 177}, 52-56.

 \refs S\'{a}nchez-Lavega A. (2004) {\em Astrophys. J., 609}, L87-90.

 \refs Sanz-Forcada J., et al. (2011) {\em Astron. Astrophys., 532}, A6.
      
 \refs Scalo J., et al. (2007) {\em Astrobiol., 7}, 85-166.

 \refs Schaefer B. E., et al. (2000) {\em Astrophys. J., 529}, 1026-1030.

 \refs Schrijver C. J., et al. (2012) {\em J. Geophys. Res., 117}, CiteID A08103.

 \refs Seager S. (2013) {\em Science, 340}, 577-581.

 \refs Seager S., et al. (2013) {\em Astrophys. J., 777}, id.95.

 \refs Segura A.,  et al. (2005) {\em Astrobiol., 5}, 706-725.

 \refs Segura A.,   et al. (2010) {\em Astrobiol., 10}, 751-771.

 \refs Sekiya M., et al. (1980) {\em Prog. Theor. Phys., 64}, 1968-1985.

 \refs Sekiya M., et al. (1981) {\em Prog. Theor. Phys., 66}, 1301-1316.

 \refs Selsis F., et al. (2008) {\em Phys. Scripta, 130}, id. 014032.

 \refs Sheeley N. R. Jr.,  et al. (1997) {\em Astrophys. J., 484}, 472-478.

 \refs Sheldon N. D. (2006) {\em Precambrian Res., 147}, 148-55.

 \refs Soderblom D. R., et al. (1993) {\em Astrophys. J., 409}, 624-634.

 \refs Stern R. A., et al. (1995) {\em Astrophys. J., 448}, 683-704.

 \refs Stewart S. T. and Leinhardt Z. M. (2012) {\em Astrophys. J., 751}, id 32.

 \refs Telleschi A., et al. (2005) {\em Astrophys. J., 622}, 653-679.

 \refs Tian F., et al. (2005) {\em ApJ, 621}, 1049-1060.

 \refs Tian F., et al. (2008) {\em J. Geophys. Res., 113}, E05008.

 \refs Tian F., et al. (2009) {\em Geophys. Res. Lett. 36}, L02205.

 \refs Tian F.,  et al. (2010), {\em  Earth Plan. Sci. Lett., 295}, 412-418.

 \refs Tsiganis K., et al. (2005) {\em Nature, 435}, 459-461.

 \refs Ueno Y., et al. (2009)  {\em Proc. Nat. Acad. Sci., 106,} 14784-14789.

 \refs Valley J. W., et al. (2002) {\em Geology, 30}, 351-354.

 \refs van den Oord G. H. J. and Doyle J. G. (1997)  {\em Astron. Astrophys., 319}, 578-588.

 \refs Vasquez M., et al. (2013a) {\em Astron. Astrophys., 549}, id.A26.

 \refs  Vasquez M., et al. (2013b) {\em Astron. Astrophys., 557}, id.A46.

 \refs Vidal-Madjar A.,  et al., (2003) {\em Nature, 422}, 143-146.

 \refs Volkov A. N. and Johnson R. E.( 2013) {\em Astrophys. J., 765},  id90.

 \refs von Paris P.,  et al. (2008) {\em Planet. Space Sci., 45}, 1254-1259.

 \refs von Paris P., et al. (2010) {\em Astron. Astrophys., 522}, id.A23.

 \refs von Paris P., et al. (2013) {\em Planet. Space Sci., 82}, 149-154.

 \refs von Zahn U., et al. (1980) {\em J. Geophys. Res., 85}, 7829-7840.

 \refs Walker J. C. G. (1977) {\em Evolution of the atmosphere}. McMillan, New York.

 \refs Walker J. C. G., et al. (1981)   {\em J. Geophys. Res., 86}, 9776-9782.

 \refs Walsh  K. J.,  et al. (2011) {\em Nature, 475}, 206-209.

 \refs Wargelin B. J. and Drake J. J. (2001) {\em Astrophys. J. Lett., 546}, L57-L60.

 \refs Watson A. J., et al.  (1981) {\em Icarus, 48}, 150-166.

 \refs Whitmire D. P., et al. (1995) {\em J. Geophys. Res., 100}, 5457-5464.

 \refs Wood B. E., et al. (2002)  {\em Astrophys. J., 574}, 412-425.

 \refs Wood B. E.,  et al. (2005). {\em Astrophys. J., 628}, L143-146.

 \refs Wordsworth R. and Pierrehumbert R. (2013) {\em Science, 339}, 64-67

 \refs Wordsworth R. D., et al. (2011) {\em Astrophys. J., 733}, id.L48.

 \refs Wordsworth R. D., et al. (2010) {\em Astron. Astrophys., 522}, id.A22.

 \refs Wuchterl G. (1993) {\em Icarus, 106}, 323-334.

 \refs Yelle R. V. (2004) {\em Icarus, 170}, 167-179.

 \refs Yung Y. L. and DeMore W. B. (1999) {\em Photochemisty of planetary atmospheres}, Oxford University Press, Oxford.

 \refs Zahnle K. J. and  Kasting J. F. (1986) {\em Icarus, 680}, 462-480.

 \refs Zahnle K. J., et al. (1988) {\em Icarus, 74}, 62-97.

 \refs Zahnle K., et al. (2007) {\em Space Sci. Rev., 129}, 35-78.

 \refs Zsom A., et al. (2012)  {\em Icarus, 221}, 603-616.

}

\end{document}